\documentclass[aps,pra,twocolumn,floatfix]{revtex4-1}

\usepackage{amsmath,amssymb}
\usepackage{graphicx}
\usepackage{hyperref}
\usepackage[usenames, dvipsnames]{color}

\newcommand{\wT}{\omega_T}
\newcommand{\dwT}{\delta\omega_T}
\newcommand{\dc}{\mathrm{dc}}
\newcommand{\ac}{\mathrm{ac}}
\newcommand{\phidc}{\phi_\dc}
\newcommand{\phiac}{\phi_\ac}

\newcommand{\wmod}{\omega_m}
\newcommand{\phimod}{\theta_m}
\newcommand{\wF}{\boldsymbol{\omega}}
\newcommand{\dwF}{\delta\boldsymbol{\omega}}
\newcommand{\Gammaphi}{\Gamma_\phi}

\newcommand{\pink}{1/f}
\newcommand{\white}{\mathrm{w}}
\newcommand{\sinc}{\mathrm{sinc}}
\newcommand{\dwFdc}[1]{\boldsymbol{\nu}_{\dc,#1}}
\newcommand{\dwFac}[1]{\boldsymbol{\nu}_{\ac,#1}}
\newcommand{\wir}{\omega_\mathrm{ir}}
\newcommand{\wuv}{\omega_\mathrm{uv}}
\newcommand{\Phidc}{\Phi_\dc}
\newcommand{\Phiac}{\Phi_\ac}
\newcommand{\dPhi}{\delta\Phi}
\newcommand{\dPhidc}{\dPhi_\dc}
\newcommand{\dPhiac}{\dPhi_\ac}
\newcommand{\Phiext}{\Phi}
\newcommand{\cor}{\gamma_{\phi}}
\newcommand{\Phisweet}{\Phiac^*}
\newcommand{\DCsweet}{DC sweet spot}
\newcommand{\ACsweet}{AC sweet spot}
\newcommand{\DCsweets}{DC sweet spots}
\newcommand{\ACsweets}{AC sweet spots}
\newcommand{\id}{\mathrm{I}}
\newcommand{\back}{\mathrm{bkgd}}

\newcommand{\Exp}[1]{e^{#1}}

\newcommand{\wpump}{\omega_m}

\newcommand{\eff}{\mathrm{eff}}

\newcommand{\BJ}{\mathrm{J}}

\newcommand{\units}[1]{\,\mathrm{#1}}
\newcommand{\wmean}{\overline{\omega}_T}
\newcommand{\etamean}{\overline{\eta}_T}

\begin{document}

\title{AC flux sweet spots in parametrically-modulated superconducting qubits}

\author{Nicolas Didier}
\thanks{These authors contributed equally to this paper.}
\affiliation{Rigetti Computing, 2919 Seventh Street, Berkeley, CA 94710}
\author{Eyob A. Sete}
\thanks{These authors contributed equally to this paper.}
\affiliation{Rigetti Computing, 2919 Seventh Street, Berkeley, CA 94710}
\author{Joshua Combes}
\affiliation{Rigetti Computing, 2919 Seventh Street, Berkeley, CA 94710}
\author{Marcus P. da Silva}
\affiliation{Rigetti Computing, 2919 Seventh Street, Berkeley, CA 94710}

\date{\today}

\begin{abstract}
The ubiquitous presence of $1/f$ flux noise was a significant barrier to
long-coherence in superconducting qubits until the development of
qubits that could operate in static, flux noise insensitive configurations
commonly referred to as ``sweet-spots''. Several proposals for entangling gates
in superconducting qubits tune the flux bias away from these spots, thus reintroducing 
the dephasing problem to varying degrees. Here we revisit one such
proposal, where interactions are parametrically activated by rapidly modulating the flux bias of the
qubits around these sweet-spots, and study the effect of modulation on the sensitivity to flux noise. 
We explicitly calculate how dephasing rates depend on
different components of the flux-noise spectrum, and show
that, although the qubits are parked at flux insensitive points, the modulation results in increased dephasing 
rate due to both the multiplicative $1/f$ and white noise components. 
Remarkably, we find a novel sweet spot under flux modulation, which we dub the
{\em AC sweet spot}, that is insensitive to $1/f$ flux noise.
We show that simple filtering of the flux
control signal additionally protects parametric entangling gates from
white noise in the control electronics at this AC sweet spot, allowing for interactions 
of quality that is limited only by higher order effects and other sources of noise.
\end{abstract}

\maketitle

\section{Introduction}

Low-frequency flux noise is often the limiting factor in the coherence
times of flux-tunable qubits
~\cite{Koch1983,Wellstood87,Devoret2002,Martinis2003,Ithier05,
  Yoshihara_2006,Bialczak_2007,RHKoch_2007,Faoro_2008,Manucharyan_2009,
  xmon,Martinis_2014,Wang15,Omalley15,Kumar16,Yan_2016,Quintana_2017,
  Kou_2017,Plourde17,Koch19}. Early
measurements on superconducting quantum interference devices (SQUIDs) 
showed the existence of flux noise with a
$1/f$-like noise power spectrum~\cite{Koch1983} and has been
extensively studied since then. Flux noise has been found to be universal
with a magnitude of a few 
$\mu\Phi_0$
at $1\units{Hz}$ despite differences in device design, materials and
sample dimensions~\cite{Wellstood87}. A major breakthrough in the
design of superconducting qubits was the realization that 
the impact of the $1/f$ noise depends not only on its strength, but
also by how sensitive the qubit frequency is to flux. To leading order, the
dephasing rate is proportional to the gradient of the qubit frequency
with respect to flux, resulting in DC flux bias ``sweet spots'' that
are insensitive to flux noise~\cite{Devoret2002}. This same insight
has been used to simply {\em reduce} the leading order contribution of
flux noise to dephasing (instead of eliminating it) using weakly tunable qubits~\cite{Plourde17}.

In this work, we investigate how the flux noise contributes to the
dephasing of superconducting qubits under flux modulation. It
has been shown that, even with qubits operating at first-order flux
sweet spots, the dephasing time is substantially reduced during
modulation, ultimately limiting the fidelity of
parametrically-activated entangling
gates~\cite{Rigetti-blue_2017,Rigetti-white_2017}.  This is
intuitively explained by the fact that the parametric modulation
induces frequency excursion, causing the qubit to periodically
explore regions of higher sensitivity to flux noise. Here we give a
full analytic accounting of how different types of flux noise
contribute to dephasing, focusing on broadband white noise
as well as additive and multiplicative $1/f$ noise.

Consistent with previous related work~\cite{Didier15}, we find these gates to
be first-order insensitive to additive $1/f$ flux noise, which is
particularly encouraging due to the universality of this type of
noise and the difficulty in reducing its magnitude. We show that multiplicative flux noise,
on the other hand, is more damaging, and operating points that are insensitive to 
additive $1/f$ noise do not guarantee insensitivity to multiplicative flux noise.
Despite this challenge, we present an operating point that is first-order 
insensitive to multiplicative $1/f$ flux noise---we refer this operating point as the \ACsweet{}~\footnote{Similar 
dynamical sweet spots have been previously described for charge noise in 
superconducting qubits~\cite{Sillanpaa2012}.}.
With these results, the remaining dominant source of dephasing under modulation
is the white (broadband) component of the noise. Remarkably, we
show that the impact of this source of noise can be greatly reduced by
filtering the flux line control signal, leading to a general \ACsweet{} for
tunable superconducting qubits under flux modulation---an operating
point for parametric gates that is insensitive to $1/f$ and white flux noise. 

As an application of these results, we calculated the fidelity of
parametrically activated controlled-Z gate between capacitively
coupled fixed- and tunable-frequency transmons. We find that when the
gates are operated at the \ACsweet{} with filtered flux control signal, 
one can achieve the error rates limited by the energy
relaxation times and background dephasing rates of the qubits (set by
other sources of noise, such as critical current fluctuations and thermal noise). Such \ACsweet{} has been observed experimentally, resulting in a controlled-Z gate with fidelity higher than $99\%$~\cite{Hong19}.

The outline of the paper is as follows.  In Sec.~\ref{section_rate} we
characterize the dephasing rate in the presence of additive and
multiplicative flux noise with $1/f$ and white noise statistics.  In
Sec.~\ref{section_sweet} we show how the \ACsweets{} emerge at
particular modulation amplitudes.  In Sec.~\ref{section_numerics} we
obtain the dephasing rate by numerically averaging the off-diagonal
element of the density matrix over realizations of flux noise.  In
Sec.~\ref{section_fidelity} we focus on parametric entangling gate
fidelity in presence of flux noise, showing that high-fidelity
two-qubit gates are obtained at the \ACsweet{} regardless of the $1/f$
strength.  Analytic derivations of the dephasing rate are provided in
the Appendices.

\section{Dephasing rate under modulation}
\label{section_rate}

We begin by reviewing the derivation of the dephasing rate without modulation for transmon qubits.
The transition frequencies of tunable superconducting qubits, $\wT$, can be `tuned' in time. 
This tunability is achieved by controlling the magnetic flux $\Phiext$ threading the SQUID loop.
Consequently, any noise on the magnetic flux causes fluctuations in the transition frequencies, $\dwT$, which results in dephasing.

To determine the dephasing rate we calculate the decay rates of the off-diagonal components of $\rho(t)$ evolving under the 
Hamiltonian $H(t) = [\wT(t) + \dwT(t)] |1\rangle\langle 1 |$. 
This is accomplished by transforming the Hamiltonian to an interaction picture, to remove the deterministic dynamical phase $\int_{0}^t \mathrm{d}t' \wT(t')$, and then solving the equation  $\dot{\rho} = -i[H(t),\rho]$. 

Typical dephasing rates are obtained by averaging the off-diagonal density matrix element $\rho_{01}=\langle 0|\rho|1\rangle$ over the flux fluctuations,
\begin{align}
\rho_{01}(t) = \left\langle e^{i\int_{0}^t \mathrm{d}t' \dwT(t')} \right\rangle \rho_{01}(0)
\equiv e^{-\cor(t)}\rho_{01}(0),
\label{rho01}
\end{align}
where $\langle . \rangle$ denotes the expectation over the flux fluctuations.

For Gaussian noise, for example for Gaussian $1/f$ or  white noise, one can explicitly compute the expectation at the level of the argument, i.e., $\langle \exp[i\delta \upsilon(t)]\rangle=\exp[-\frac{1}{2}\langle \delta \upsilon^2(t)\rangle]$~\cite{Sete2017}. Under reasonable physical assumptions, such as adiabaticity of evolution under low frequency flux noise, it can be shown that the predictions from Gaussian noise are valid~\cite{PGFA14}. We may define the dephasing rate due to these noise sources, $\Gammaphi$, via
\begin{align}
\cor(t)&=\tfrac{1}{2}\int_{0}^t\mathrm{d}t_1\int_{0}^t\mathrm{d}t_2 \langle \dwT(t_1)\dwT(t_2)\rangle \equiv (\Gammaphi t)^\beta.
\label{correlationdef}
\end{align}
This expression shows that the dephasing rate can be extracted from the time evolution of $\cor(t)$ by fitting to a power law. 
The exponent $\beta$ depends on the noise statistics and is usually between 1 and 2.

To control the transition frequency $\omega_T(t)$, so as to activate parametric entangling gates, we must control the magnetic flux $\Phi(t)$, and for the purposes of the discussion here, we consider sinusoidal modulation taking the form
\begin{align}
\Phiext(t)  = \Phidc
+ \Phiac\cos(\wmod t+\phimod),
\end{align}
where the controllable parameters are: a dc offset in the flux $\Phidc$ -- called the {\em parking flux}, the amplitude of the ac flux $\Phiac$, the modulation frequency $\wmod$, and the 
modulation phase $\phimod$. Given both the parking flux and the modulation amplitude are subjected to noise, $\dPhidc(t)$ and $\dPhiac(t)$ respectively, the fluctuating flux bias is  
\begin{align}
\Phiext(t) +\delta\Phi(t) = \Phidc&+\dPhidc(t) \nonumber\\
+ [\Phiac&+\dPhiac(t)]\cos(\wmod t+\phimod).
\label{totalflux}
\end{align}
The flux fluctuation $\dPhidc$ is referred to as additive noise and $\dPhiac$ as multiplicative noise.
The spectral density of the additive noise $S_\dc(\omega)$ is defined by
\begin{align}
\langle \dPhidc(t_1)\dPhidc(t_2)\rangle=\int_{-\infty}^\infty\frac{\mathrm{d}\omega}{2\pi}S_\dc(\omega)e^{i\omega(t_1-t_2)},
\end{align}
and similarly for the multiplicative noise $\dPhiac(t)$.

\begin{figure}[t]
\includegraphics[width=\columnwidth]{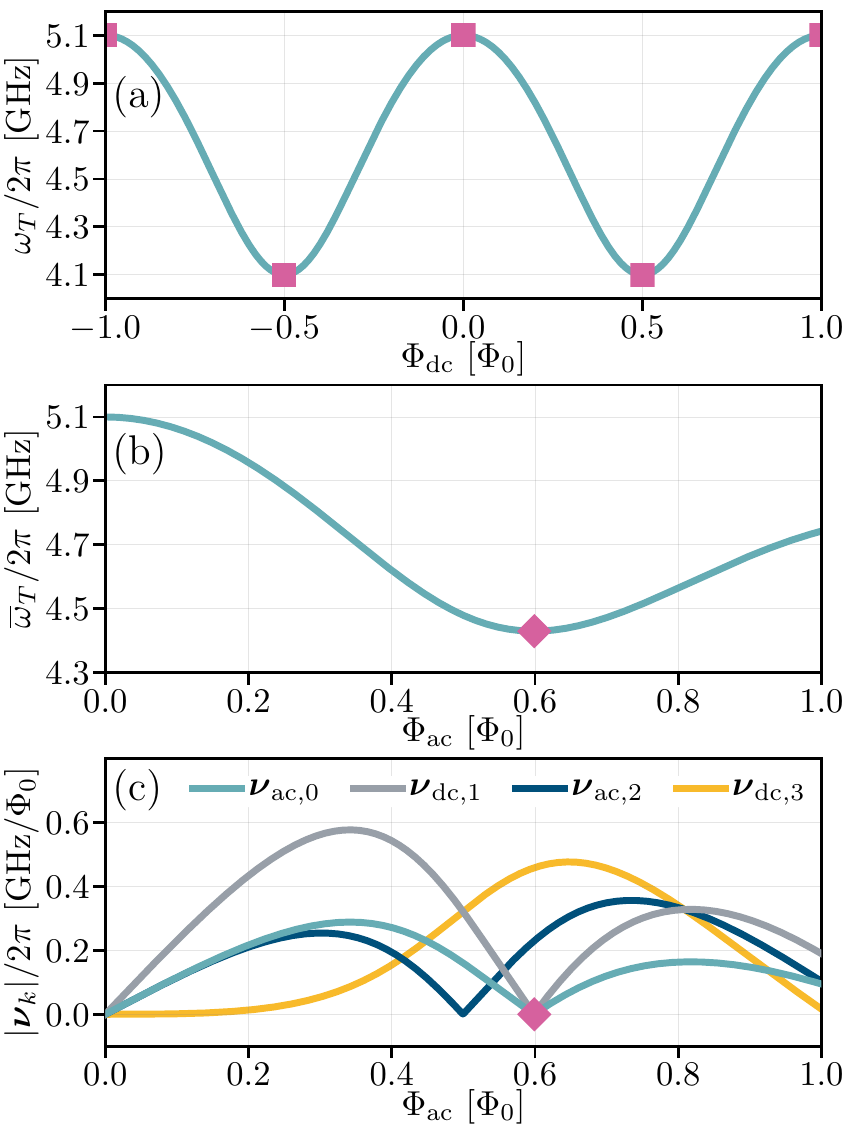}
\caption{DC and AC flux sweet spots in asymmetric transmons.
(a) Frequency as a function of DC flux bias for $\Phiac=0$, \DCsweets{} are located at the extrema of the band (squares).
(b) Average frequency as a function of the modulation amplitude for $\Phidc=0$, the \ACsweet{} is located at the minimum, $\Phisweet\approx 0.6\,\Phi_0$ (diamond).
(c) Derivatives of the Fourier series with respect to $\Phidc$ and $\Phiac$.
Both $\dwFac{0}$ and $\dwFdc{1}$ vanish at $\Phisweet$ (diamond), thereby providing immunity to $1/f$ and lowpass-filtered white flux noise.}
\label{fig_Fourier}
\end{figure}

Because the flux-to-frequency transduction is highly nonlinear in tunable superconducting qubits, the qubit frequency $\wT$ oscillates at many harmonics of the modulation frequency.
A convenient representation of the qubit frequency under modulation is its Fourier series~\cite{Nico17}.
The Fourier coefficients $\wF_k$ being dependent on the flux pulse parameters $\Phidc$ and $\Phiac$, frequency fluctuations arise in fluctuations $\dwF_k$ of the Fourier coefficients $\wF_k$,
\begin{align}
\wT(t)+\dwT(t)&=\sum_{k=0}^\infty[\wF_k+\dwF_k(t)]\cos[k(\wmod t+\phimod)].
\end{align}
For small noise amplitude, the leading contribution of flux noise in frequency fluctuations is via its slope, $\partial\wT/\partial\Phi$. When the first derivative vanishes the resulting dephasing time from the second derivative is well above the limits imposed by other noise sources (at least for typical transmon parameters), and therefore, for simplicity, we disregard the effects of the second derivative. In the Fourier space, frequency fluctuations are equal to,
\begin{align}
\dwF_k(t)=\frac{\partial\wF_k}{\partial\Phidc}\dPhidc(t)+\frac{\partial\wF_k}{\partial\Phiac}\dPhiac(t).
\end{align}
The Fourier coefficients as well as their derivatives are provided in Appendix~\ref{appendix_Fourier}.
In the long time limit, the dephasing rate is found from,
\begin{align}
\cor(t)&=t^2\int_{\wir}^{\wuv}\frac{\mathrm{d}\omega}{2\pi}
\sinc^2(\tfrac{1}{2}\omega t)
\left[\dwFdc{0}^2S_\dc(\omega)+\dwFac{0}^2S_\ac(\omega)\right]\nonumber\\
&+t\sum_{k=1}^{k_\mathrm{uv}}\tfrac{1}{4}\left[\dwFdc{k}^2S_\dc(k\wmod)+\dwFac{k}^2S_\ac(k\wmod)\right],
\label{simplecor}
\end{align}
with the shorthand notation 
$\dwFdc{k}=\partial\wF_k/\partial\Phi_\dc$ and 
$\dwFac{k}=\partial\wF_k/\partial\Phi_\ac$, see Appendix~\ref{appendix_rate}.
The expression of $\cor(t)$ in Eq.~\eqref{simplecor} highlights an important property of dephasing under flux modulation: the dephasing rate is governed by the noise spectrum around harmonics of the modulation frequency, the qubit hence probes the environment at $k\wmod$.
The frequencies $\wir$ and $\wuv$ are the infrared and ultraviolet cutoffs, corresponding to the longest and shortest timescales of the experiment, and $k_\mathrm{uv}$ is the index of the last harmonic below the cutoff.
Equation~\eqref{simplecor} provides the leading term of $\cor(t)$ used to define the dephasing rates, there are however additional oscillations at harmonics of the modulation frequency with small amplitudes, scaling as $\propto1/\wmod$ (see Appendix~\ref{appendix_rate}).
The first derivatives $\dwFdc{k},\dwFac{k}$ are plotted in Fig.~\ref{fig_Fourier}(c).

The flux noise can have low-frequency and/or high-frequency components. 
In the following, we address low-frequency $1/f$ noise (also referred to as pink noise) and white noise as the leading mechanisms for dephasing under modulation. The total spectral density on the DC flux bias can be written as
\begin{align}
S_\dc(\omega)&=\frac{2\pi}{|\omega|}A_{\dc,\pink}^2 + A_{\dc,\white}^2,
\end{align}
where $A_{\dc,\pink}$ and $A_{\dc,\white}$ are the  $1/f$ and white noise amplitudes, respectively.
Low frequency flux noise has been measured experimentally with a dependence $\propto|\omega|^{-\alpha}$ with $\alpha\approx 1$~\cite{Wellstood87,Omalley15,Plourde17}, here we consider $\alpha=1$ for simplicity.
Similar definition for the spectral density of the noise on the AC flux bias, $S_\ac(\omega)=\frac{2\pi}{|\omega|^{\alpha}}A_{\ac,\pink}^2 + A_{\ac,\white}^2$. As seen from the structure of Eq.~\eqref{simplecor}, $1/f$ and white flux noise contributes qualitatively differently to the dephasing rate.
First, because $1/f$ has the largest magnitude at low-frequencies, the contributions at $\wmod$ and higher harmonics vanish;
$1/f$ flux noise thus only depends on the fundamental harmonic ($k=0$) $\wF_0$, i.e., the average qubit frequency under modulation
\begin{align}
\wF_0=\wmean = \frac{1}{T}\int_{0}^{T}\omega_T(t')dt',
\end{align}
where $T = 2\pi/\wpump$ is the modulation period. Second, the remaining integral over frequency in Eq.~\eqref{simplecor} has different scalings with respect to time:
quadratic for $1/f$ noise and linear for white noise.
The dephasing function finally reads 
\begin{align}
\cor(t) = (\Gamma_{\phi,\pink} t)^2+\Gamma_{\phi,\white} t,
\label{dephasing_function}
\end{align}
where the dephasing rates are 
\begin{align}
\Gamma_{\phi,\pink}&=\lambda\sqrt{\dwFdc{0}^2A_{\dc,\pink}^2+\dwFac{0}^2A_{\ac,\pink}^2},\label{Gammapink}\\
\Gamma_{\phi,\white}&=\frac{1}{4}\sum_{k=0}^{k_\mathrm{uv}}(1+\delta_k)\left[\dwFdc{k}^2A_{\dc,\white}^2+\dwFac{k}^2A_{\ac,\white}^2\right].\label{Gammawhite}
\end{align}
In the dephasing rate due to $1/f$ noise, the parameter
$\lambda=\sqrt{\ln(e^{3/2-\gamma}/\wir t)}$ (see Appendix~\ref{appendix_rate})
is typically $\lambda\approx3$~\cite{Plourde17} with $\gamma$ the Euler constant, $\delta_k=1$ if $k=0$ and $0$ otherwise.
The dephasing rates described in Eq.~\eqref{dephasing_function} have a lot of structure that is not immediately apparent, but which will be described in the next section.

\section{AC flux sweet spots}
\label{section_sweet}
In order to minimize the effect of flux noise, the tunable qubit is parked at a \DCsweet{} [see Fig.~\ref{fig_Fourier}(a)].
We consider $\Phidc=0$ in the following for simplicity.
The qubit frequency is symmetric around such parking point, as a consequence under flux-bias modulation the qubit frequency oscillates only at even harmonics of the modulation frequency, i.e.~$\wF_{2k+1}=0$.
For example, at small modulation amplitudes the frequency oscillates at twice the modulation frequency.
The derivative of the odd harmonics with respect to $\Phiac$ are also equal to zero, i.e.~$\dwFac{2k+1}=0$.
Moreover, the even harmonics are even functions of $\Phidc$, the slope vanishes at the \DCsweet{}, i.e., ~$\dwFdc{2k}=0$.
If the parking flux is slightly moved away from the \DCsweet{}, the odd harmonics $\wF_{2k+1}$ are activated but the even harmonics are not affected at first order (see Appendix~\ref{appendix_Fourier}).

A tunable qubit parked at its \DCsweet{} remains first-order insensitive to additive $1/f$ flux noise during modulation, as was found in Ref.~\onlinecite{Didier15} (the modulation acts as a dynamical decoupling for additive low-frequency noise).
Indeed, because $\dwFdc{0}=0$, the dephasing rate due to $1/f$ noise is due solely to the multiplicative low-frequency noise,
\begin{align}
\Gamma_{\phi,\pink}&=\lambda\left|\frac{\partial \wmean}{\partial \Phiac}\right|A_{\ac,\pink}.
\label{Gammaacpink}
\end{align}
This expression of dephasing rate is analogous to the usual dephasing rate under $1/f$ noise in the absence of modulation (see, e.g., Ref.~\onlinecite{Sete2017} and references therein). The major difference in presence of flux modulation is that the qubit frequency $\wT$ is replaced by the averaged qubit frequency under modulation $\wmean$.
As a consequence, the same way \DCsweets{} appear at extrema of the tunable qubit frequency under DC flux bias, \ACsweets{} emerge at extrema of the average frequency of the tunable qubit under flux modulation.
That is, if $1/f$ flux noise is dominating over white noise, the \ACsweet{} $\Phisweet$ are located at,
\begin{align}
\frac{\partial \wmean}{\partial \Phiac}(\Phisweet)=0.
\end{align}
This modulation amplitude is rather large, the first \ACsweet{} is found around 
the first zero of the Bessel function $\BJ_1$,
$\Phisweet\approx0.6\,\Phi_0$ as shown in Fig~\ref{fig_Fourier}(b).
 Note that, from the expressions of Appendix~\ref{appendix_Fourier}, \ACsweets{} are also found around $\Phidc\approx\pm\frac{1}{4}\Phi_0$ and $\Phiac\approx0.4\,\Phi_0$, here mainly set by the first zero of the Bessel function~$\BJ_0$.

In the presence of strong white flux noise, such \ACsweets{} vanish.
This is due to the contribution of high frequency noise at harmonics weighted by $\dwFdc{2k+1}$ and $\dwFac{2k}$.
Let us now consider that the additive and multiplicative white noise are filtered such that the contribution from harmonics $k\geq2$ is strongly reduced, i.e., $k_\mathrm{uv}=1$. This is obtained with a lowpass filter on $\dPhidc$ and $\dPhiac$ with a cutoff frequency between $\wmod$ and $2\wmod$ (note that while filtering the noise on the DC and AC flux signals independently may be challenging experimentally, it is instructive to consider this scenario). Then, the identity
\begin{align}
\frac{\partial \wF_1}{\partial \Phidc}=2\frac{\partial \wmean}{\partial \Phiac},
\end{align}
establishes that the dephasing rate due to white noise is proportional to $\dwFac{0}^2$, 
\begin{align}
\Gamma_{\phi,\mathrm{lp}\white}&=\left(\frac{\partial \wmean}{\partial \Phiac}\right)^2(A_{\dc,\white}^2+\tfrac{1}{2}A_{\ac,\white}^2),
\label{Gammawhitefilter}
\end{align}
and hence vanishes at $\Phisweet$, as shown in Fig.~\ref{fig_Fourier}(c).
At \ACsweets{}, tunable superconducting qubits are thus immune to $1/f$ noise and lowpass-filtered white noise.

A single lowpass filter in the shared signal path for the DC and AC signals can be used, instead of separate filters for those separate parts of the signal.
The resulting dephasing rate is equal to
$\Gamma_{\phi,\mathrm{lp}\white}=\frac{1}{4}\dwFdc{1}^2S(\wmod)$,
with $S(\omega)$ the spectral density of the total white flux noise
$\delta\Phiext(t)$.
Its value at $\wmod$, 
$S(\wmod)=A_{\dc,\white}^2+\tfrac{1}{2}A_{\ac,\white}^2$, 
combines noise components around $0\,\wmod$, $1\,\wmod$, $2\,\wmod$ and
leads to the same dephasing rate as in Eq.~\eqref{Gammawhitefilter} when both components are filtered independently.

As a conclusion, the \ACsweets{} of a tunable qubit under flux modulation subjected to $1/f$ and white flux noise are obtained by reducing or filtering white noise.
Such sweet spots are limited by white noise due to thermal noise such as coming from the coupling through the qubit readout line, and to low-frequency charge and critical current noise.
\ACsweets{} are optimal operating points for parametric entangling gates~\cite{Nico17,Rigetti-blue_2017,Rigetti-white_2017} to reduce the effect of decoherence on two-qubit gate fidelity.
High fidelity two-qubit gates are obtained by designing superconducting circuits to operate their parametric entangling gates at $\Phidc=0$ and $\Phiac=\Phisweet$.

\section{Dephasing under modulation: Numerical simulations}
\label{section_numerics}

In this section, we numerically generate $1/f$  and white flux noise \cite{Note}, and determine the dephasing rate during the flux modulation. As described in Sec.~\ref{section_rate}, the total flux  that controls the qubit frequency depends on the DC flux ($\Phidc$)  and AC flux ($\Phiac$) signals. Each signal can have low- and high-frequency noise components as described in Eq.~\eqref{totalflux}.
In the following, we compute the contributions of the additive noise $\dPhidc(t)$ and multiplicative noise $\dPhiac(t)$ for both white noise and $1/f$ noise to the dephasing rate.  We extract the dephasing rate by fitting the decay of the off-diagonal density matrix element of the qubit.

\begin{figure}[t]
    \centering
    \includegraphics[width=\columnwidth]{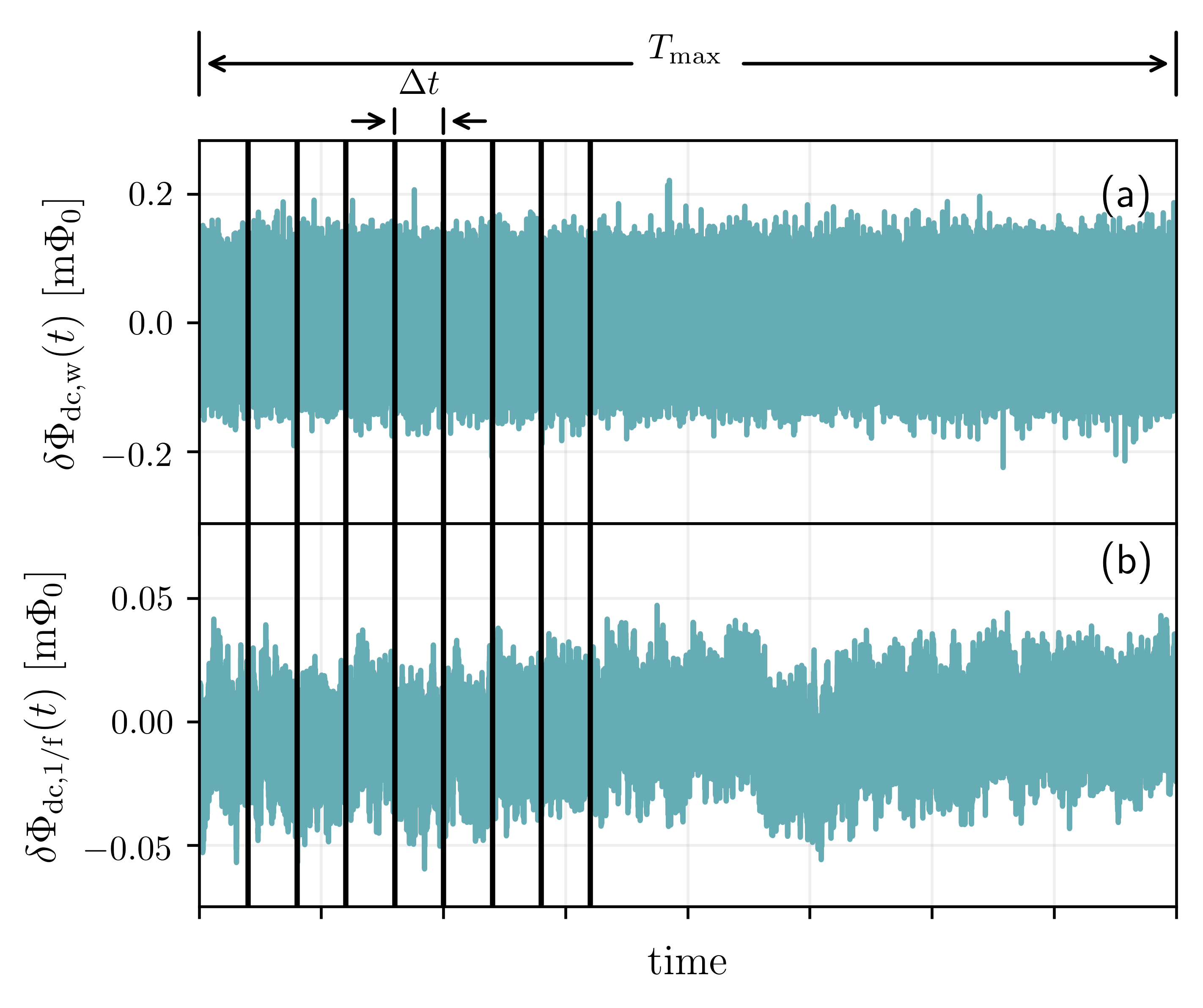}
    \includegraphics[width=\columnwidth]{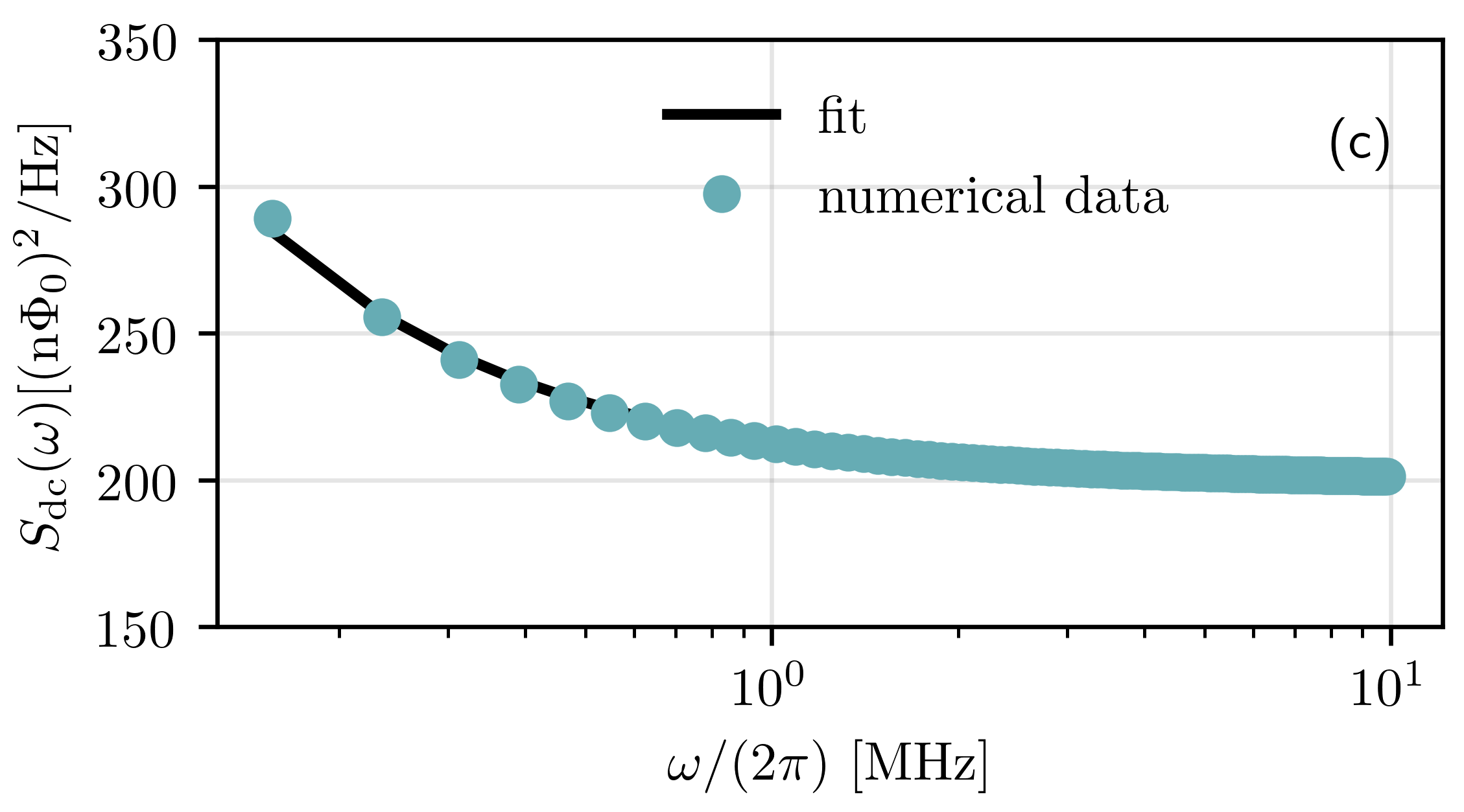}
    \caption{
    Numerically generated time traces of white noise (a) and $1/f$ noise (b). The corresponding total spectral density $S_\dc(\omega)= 2\pi\times A_{\dc,1/f}^2/|\omega|+A_{\dc,\white}^2$, where the white noise spectral density $A_{\dc,\white}= 10\,\mathrm{n}\Phi_0/\sqrt{\mathrm{Hz}}$ and $1/f$ noise amplitude $A_{\dc,\pink}= 3.63\,\mu\Phi_0$ (c).}
    \label{time_trace_spectral_density}
\end{figure}

\begin{figure}[t]
\includegraphics[width=\columnwidth]{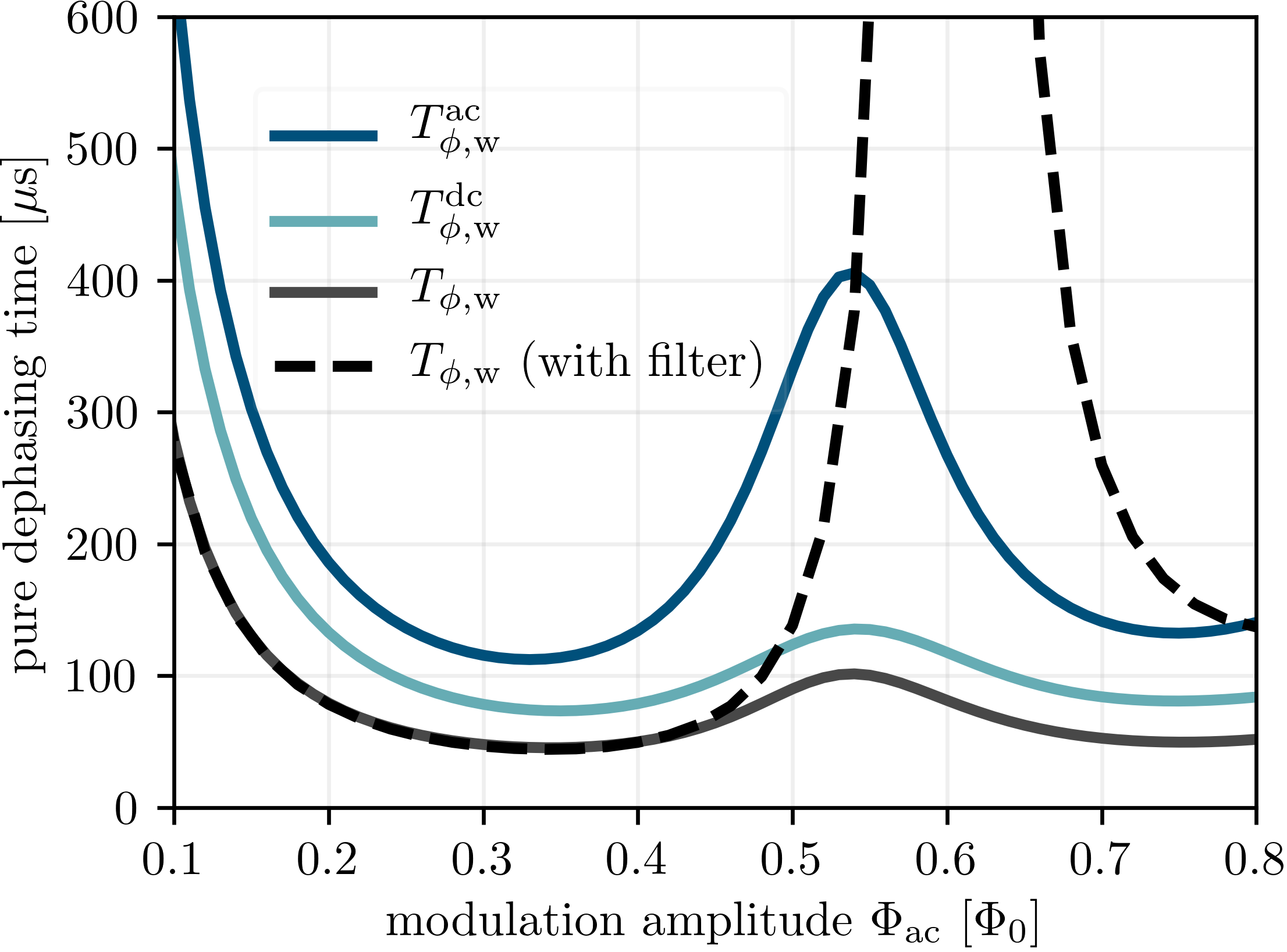}
\caption{Dephasing time during flux modulation amplitude due to multiplicative (or AC) white noise (blue), additive (or DC) white noise (teal),  both AC and DC white noise (yellow), and low-pass filter with cutoff between $\wmod$ and $2\wmod$ (dashed-black). The noise power spectral density for AC and DC white noise assumed to be $A_{\dc,\white} =A_{\ac,\white}=10\,\mathrm{n}\Phi_0/\sqrt{\mathrm{Hz}}$.
The dashed black line is the dephasing rate obtained by using a lowpass filter on the flux bias line, generating an \ACsweet.}
\label{dephasing_white}
\end{figure}

\begin{figure}[t]
\includegraphics[width=\columnwidth]{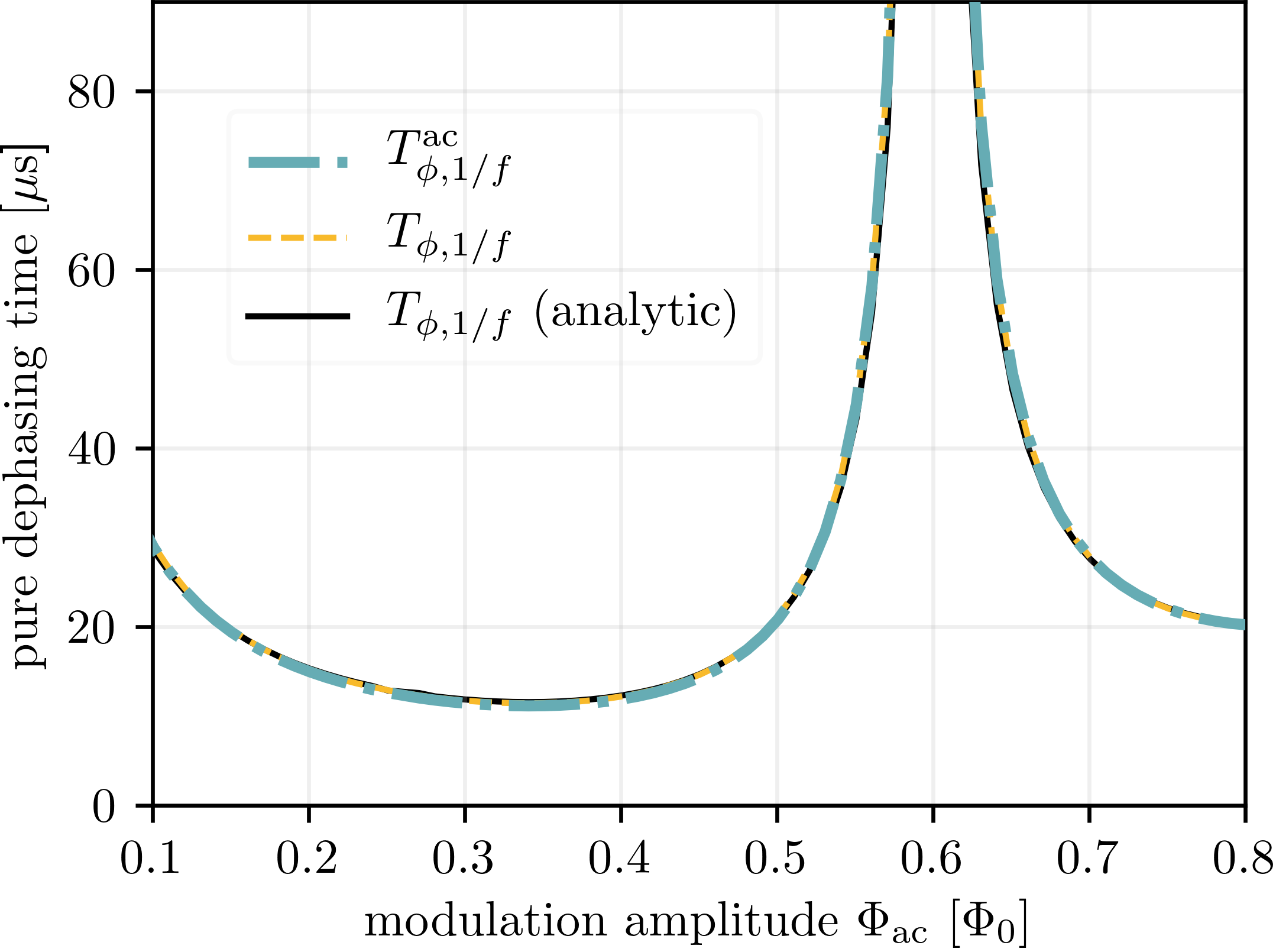}
\caption{Dephasing time during flux modulation amplitude due to multiplicative low-frequency $1/f$ flux noise (dash-dotted teal), and due to both multiplicative and additive $1/f$ flux noises (dashed yellow). Note that the dephasing time due to additive $1/f$ noise is three orders of magnitude longer than that of multiplicative $1/f$ noise (not shown). The black line is the dephasing time obtained from the analytic expression, Eq.~\eqref{Gammaacpink}. The noise power spectral density for additive and multiplicative $1/f$ flux noise are assumed to have the same strength with noise amplitude at $1\units{Hz}$, $A_{\dc,\pink} =A_{\ac,\pink}=3.63\,\mu\Phi_0$.}
\label{dephasing_one_on_f}
\end{figure}

\subsection{White flux noise}

We first consider the high-frequency additive as well as multiplicative white flux noise. These noises arise from the slow and fast flux sources and can cover a wide range of frequencies.  An example Gaussian white noise time trace is shown in Fig.~\ref{time_trace_spectral_density}(a). Here we assumed the additive and multiplicative white noise have the same power spectral density. The white noise through AC flux bias has similar time trace as the DC flux bias. 

The total length $T_{\rm max}$ of the time trace data is divided into $N$ shot events, each having a length of $dt$, where $T_{\rm max} = N dt$. The maximum frequency is determined by $1/d t$, while the lowest frequency resolution is limited by $1/T_{\rm max}$. We consider a time trace of length $T_{\rm max} = 1\units{s}$ which has $2\times 10^7$ data points. The time trace data is sliced in to $5000$ time windows or measurements each of length $\Delta t = 250\,\mu\mathrm{s}$.

For a fixed DC flux bias $\Phidc$, modulation amplitude $\Phiac$, noise amplitudes $\dPhidc$ and $\dPhiac$ (assuming only white noise), the dephasing rate is extracted from the evolution of the off-diagonal density matrix element  $\rho_{01}(t)$. Equation \eqref{rho01} is calculated for time window $\Delta t$ and repeated for all $5000$ measurements. Averaging $\rho_{01}(t)$ over all the 5000 measurements yields a decaying  off-diagonal density matrix element. Fitting the numerical data to $e^{-\Gamma_{\phi,\rm w} t}$ gives the dephasing rate $\Gamma_{\phi,\white}$. 

For the dephasing rate analysis, we use an asymmetric tunable transmon qubit with $\wT(\Phi=0)/2\pi =5.1\units{GHz}$, $\wT(\Phi=\Phi_0/2)/2\pi = 4.1\units{GHz}$, and anharmonicity  $\eta_{T}(\Phi=0)/2\pi = 0.2\units{GHz}$. We park the qubit at a \DCsweet{} ($\Phidc=0$) and vary the fast flux modulation amplitude $\Phiac$ to reveal the dependence of the pure dephasing time as function of modulation amplitude. Let us start with white noise only on the DC flux signal, i.e., the flux pulse has the form $\Phi = \dPhidc(t)+\Phiac\sin(\wmod t)$. For a spectral density strength  $A_{\dc,\white} = 10\,\mathrm{n}\Phi_0/\sqrt{\mathrm{Hz}}$, the white noise on the DC flux signal strongly contributes to dephasing rate of the qubit as shown by the teal curve in Fig.~\ref{dephasing_white}. This result agrees well with our analytic derivation in Sec.~\ref{section_rate}. If we assume a multiplicative white noise flux signal, i.e., $\Phi=[\Phiac+\dPhiac(t)]\sin(\wmod t)$, we obtain similar dephasing time behavior as a function of flux modulation amplitude (blue curve), but slightly weaker contributions to the dephasing rate. We have found that when there are both additive and multiplicative white noise at the same time, the resulting dephasing rate is the sum of the individual contributions (gray curve)
$\Gamma_{\phi,\white} = \Gamma_{\phi,\white}^{\dc}+\Gamma_{\phi,\white}^{\ac}$,
where $\Gamma_{\phi,\white}^{\dc}$ and $\Gamma_{\phi,\white}^{\ac}$ are the dephasing rates due to additive and multiplicative white noise, respectively. When low-pass filter with cutoff $\sim 1.5 \,\wmod$ is applied (dashed-black curve in Fig. \ref{dephasing_white}), one can recover the \ACsweet{} as predicted by Eq.~\eqref{Gammawhitefilter}.

\subsection{Low-frequency 1/f flux noise}

Here we analyze the contribution of low-frequency $1/f$ flux noise on the DC and AC signals to the dephasing rate during modulation. Following the same line of reasoning, we first numerically generate a long time trace of noise signal that yields $1/f$ noise power spectral density of the form $A_{\dc,\pink}^2/|f|^{\alpha}$, with $\alpha\approx 1$ (here $f=\omega/2\pi$). For simplicity we use $\alpha=1$ in this paper. The time trace is sliced to $N$ measurements each of length $\Delta t$. An example time trace of $1/f$ noise is shown in Fig. \ref{time_trace_spectral_density}(b) with a power spectral density [Fig. \ref{time_trace_spectral_density}(c)] of amplitude $A_{\dc,\pink} = 3.63\,\mu\Phi_0$. We then integrate Eq.~\eqref{rho01} for each shot and average the trajectories over the number of measurements for a given modulation amplitude $\Phiac$ and DC flux bias $\Phidc$. Fitting the numerical data to a Gaussian decay function $e^{-(\Gamma_{\phi} t)^{\beta}}$, we extract the dephasing rate $\Gamma_{\phi}$ and the exponent $\beta\approx2$. 
We repeat this procedure for different values of the modulation amplitude. 

We first consider low-frequency pure $1/f$ flux noise on the DC flux bias, i.e., $\Phi=\Phidc+\dPhidc(t)+\Phiac\sin(\wmod t)$. To minimize the low-frequency noise, the qubits are parked at \DCsweet{} $\Phidc=0$. We found that the qubit remains first-order insensitive (only limited by the second order sensitivity) to the low-frequency $1/f$ DC flux noise during modulation with dephasing time in milliseconds. This confirms the result reported earlier~\cite{Didier15} and our analytic derivation.

The other possibility is that the amplitude of the AC flux signal may be susceptible to low-frequency $1/f$ flux noise. For the same noise level as the noise on DC flux signal, the multiplicative low-frequency $1/f$ flux noise has significant contributions to dephasing [see Fig. \ref{dephasing_one_on_f} (teal curve)]. As mentioned in the previous sections, the dephasing rate due to multiplicative $1/f$ flux noise follows the gradient of the average frequency as a function of flux modulation amplitude with the \ACsweet{} appearing at the minimum of the average frequency [see Fig. \ref{dephasing_one_on_f}]. 

As can be seen in Fig.~\ref{dephasing_one_on_f} the analytic dephasing rate Eq.~\eqref{Gammaacpink} agrees well with numerical result. To get the total dephasing rate due to low-frequency $1/f$ flux noise, we added the time traces of the DC and AC noise signals to the flux bias as in Eq.~\eqref{totalflux} and fitted the decaying off-diagonal element to $e^{-(\Gamma_{\phi} t)^{\beta}}$. The extracted dephasing rate agrees with the total rate for uncorrelated additive and multiplicative noises, Eq.~\eqref{Gammapink}.

\section{Two-qubit gate fidelity }
\label{section_fidelity}

As an application of the results obtained in the previous sections, we compute the fidelity of a parametrically-activated controlled-Z gate between capacitively coupled fixed- and tunable-frequency transmons. In particular, we shed light on the importance of operating the gate at \ACsweets{} to maximize the performance of the gate. This holds irrespective of the strength of the multiplicative $1/f$ flux noise strength as long as the white noise level through the flux bias line and background noise are sufficiently weak or if the white noise is appropriately filtered. Note that in addition to choosing an operating point where the dephasing time is long, the performance of parametrically activated gates can be enhanced by 
optimizing the Hamiltonian for low coherent errors and short gate time at the \ACsweet{}.
In the analysis below, we use a Hamiltonian designed to satisfy this requirement.

For the numerical study, we model the transmon qubit as an anharmonic oscillator truncated to three levels.
The Hamiltonian of the coupled system is given by
\begin{align}
H &= \omega_{F}|1\rangle\langle 1|\otimes \id +(2\omega_{F}-\eta_F)|2\rangle\langle 2|\otimes \id\notag\\
&+\omega_{T}(t) \id\otimes|1\rangle\langle 1 |+[2\omega_{T}(t)-\eta_T(t)] \id\otimes|2\rangle \langle 2| \notag\\ 
&+g(\sigma_F^{\dag}\sigma_{T}+\sigma_{F}\sigma_{T}^{\dag}+\sigma_{F}\sigma_{T}+\sigma_{F}^{\dag}\sigma_{T}^{\dag})
\end{align}
where $\omega_F$ and $\omega_T$ are the transition frequencies of the fixed- and tunable-frequency transmons, and $\eta_F$ and $\eta_T$ are their corresponding anharmonicities, respectively; $g$ is the capacitive static coupling between the two qubits,  $\sigma_{F} = (|0\rangle\langle 1|+\sqrt{2}|1\rangle\langle 2|)\otimes \id$ and $\sigma_{T}= \id\otimes(|0\rangle\langle 1|+\sqrt{2}|1\rangle\langle 2|)$ are the lowering operators for fixed and tunable qubits with $\id$ the identity operator. 
The qubit frequency and anharmonicity are obtained from perturbation theory to $10^\mathrm{th}$ order in the small parameter $\xi=\sqrt{2E_C/E_J}$~\cite{Nico17}.
Under flux modulation, the parameter evolves as 
$\xi(t)=\sqrt{2E_C/E_{J_\eff}(t)}$ 
with the effective Josephson energy equal to 
$E_{J_{\rm eff}}(t) = \sqrt{E_{J1}^2+E_{J2}^2+2E_{J1}E_{J2}\cos[2\pi \Phi(t)/\Phi_0]}$. The flux bias is similar to Eq.~\eqref{totalflux} but with a pulse with smooth rising and falling edges to account for finite control bandwidth

\begin{align}
\Phiac(t) =\frac{\Phiac}{2} \left[\mathrm{erf}\left(\frac{t-t_{\rm ramp}}{\sigma}\right)-\mathrm{erf}\left(\frac{t+t_{\rm ramp}-t_f}{\sigma}\right)\right],
\end{align}
where $t_{\rm ramp}$ is the pulse rise time, $t_f$ is the total length of the pulse, and 
$\sigma=t_{\rm ramp}/4\sqrt{2 \ln(2)}$. 

The decoherence effects are introduced in the fidelity calculation by numerically solving a phenomenological quantum master equation
\begin{align}
\dot\rho &= -i[H,\rho]+\Gamma_{1,F}\mathcal{D}[\sigma_{F}]\rho+\Gamma_{1,T}(\Phiac)\mathcal{D}[\sigma_{T}]\rho\notag\\
&+2\Gamma_{\phi,F} \mathcal{D}[\sigma^{\dag}_{F}\sigma_{F}]\rho+2\Gamma_{\phi,\white}(\Phiac)\mathcal{D}[\sigma^{\dag}_{T}\sigma_{T}]\rho\notag\\
&+2\beta t^{\beta-1}\Gamma_{\phi,\pink}^\beta(\Phiac) \mathcal{D}[\sigma_T^{\dag}\sigma_T]\rho\notag\\
&+2\Gamma_{\phi,\back}\mathcal{D}[\sigma^{\dag}_{T}\sigma_{T}]\rho,
\label{tdQME}
\end{align}
where $\mathcal{D}[c]\rho=c\rho c^{\dag}-(c^{\dag}c\rho+\rho c^{\dag}c)/2$, $\Gamma_{1,F}$ and $\Gamma_{1,T}$ are the fixed- and tunable-frequency $|1\rangle\rightarrow |0\rangle$ energy relaxation rates, respectively,  $\Gamma_{\phi,F}$ is pure dephasing rate for fixed qubit, and $\Gamma_{\phi,\white}$ is the pure dephasing rates due to Gaussian white noise. $\Gamma_{\phi,\pink}$ is the pure dephasing rate due to low-frequency $1/f$ noise and $\Gamma_{\phi,\back}$ is the background dephasing rate, which is independent of flux bias. 
The time-dependent Lindbladians generate the appropriate decay for $1/f$ flux noise.
As shown in Appendix~\ref{appendix_QME}, these superoperators provide an average process fidelity that is slightly underestimated with respect to averaging the dynamics over $1/f$ flux noise.
Note that not only the dephasing rate but also the energy relaxation rate for tunable qubit  $\Gamma_{1,T}$ varies in modulation amplitude because of the Purcell effect~\cite{Purcell,Houck2008,Sete2014}. An alternative way to include all decoherence effects (Markovian and non-Markovian decays) in the fidelity calculation is by averaging the Markovian master equation (including decay and dephasing due to white noise) over the distribution of the $1/f$ flux noise \cite{DiCarlo19}.

\begin{figure}[t]
	\centering
	\includegraphics[width=\columnwidth]{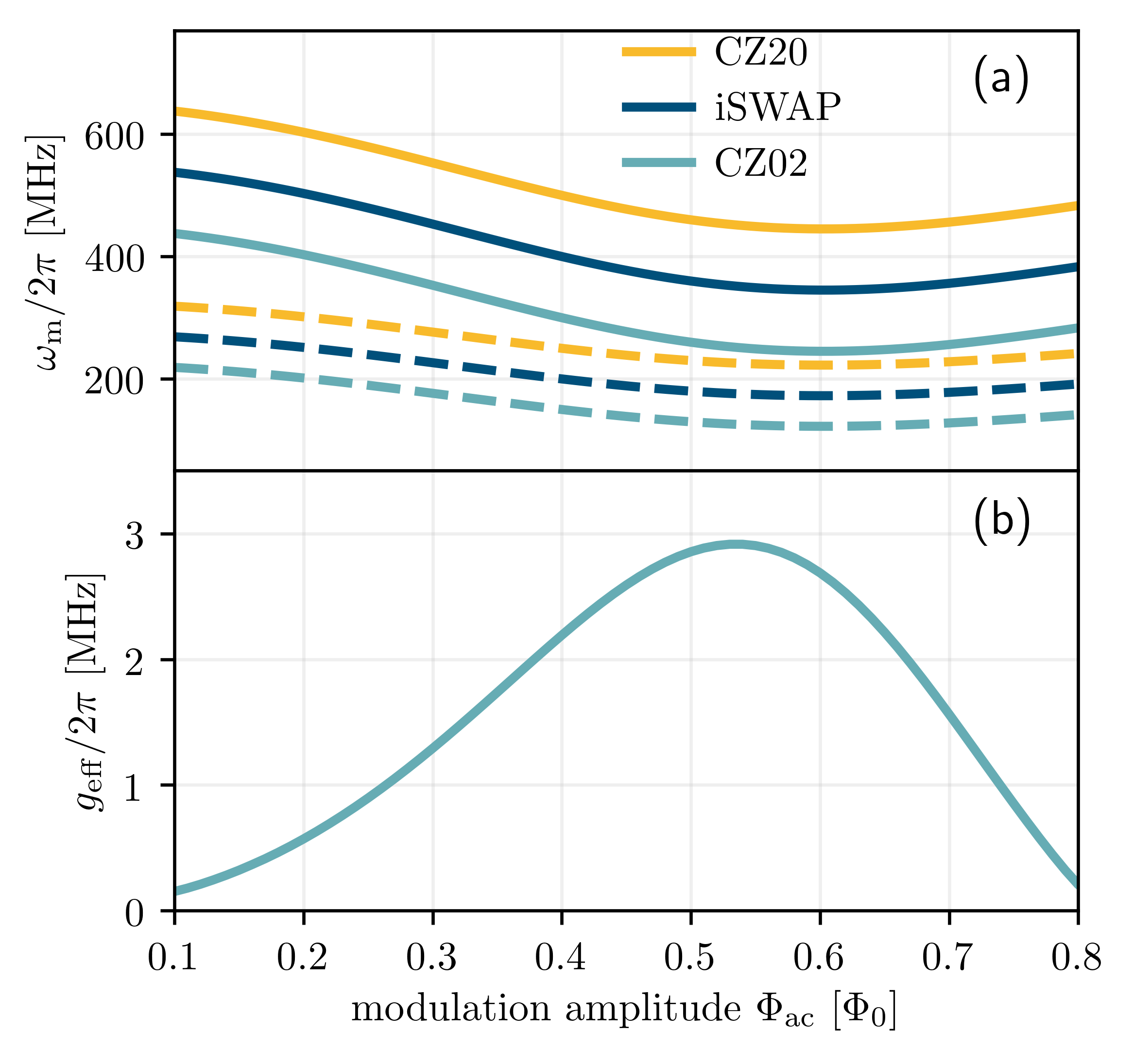}
	\caption{(a) Flux modulation frequency for activating controlled-Z gates: CZ02 (solid-teal) and CZ20 (solid-yellow), and an iSWAP gate (solid-blue) vs flux modulation amplitude $\Phi_{\rm ac}$. The dashed curves are the corresponding second harmonic resonances generated by the modulation. (b) The effective coupling $g_{\rm eff}$ between $|11\rangle$ and $|02\rangle$ states (in $|ij\rangle$ the index $i$ is for fixed- and $j$ is for tunable-frequency qubit.) }
	\label{modulation_frequency}
\end{figure}

\begin{figure}[t]
	\centering
	\includegraphics[width=\columnwidth]{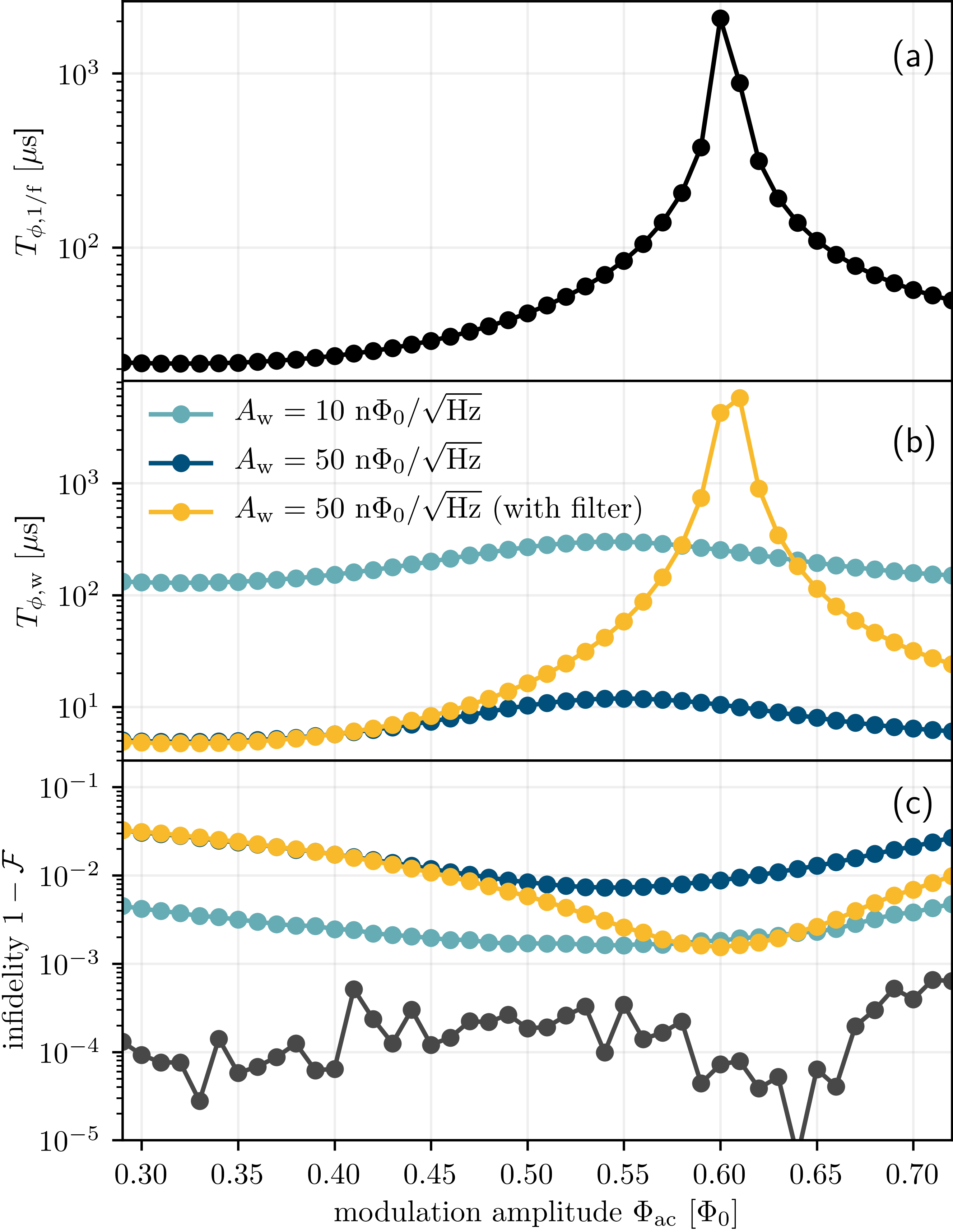}
	\caption{
	(a) Dephasing time vs flux modulation amplitude due to $1/f$ (additive and multiplicative) noise only for noise amplitude $A_{\dc,\pink}=A_{\ac,\pink} =3.63\,\mu\Phi_{0}$. Note that the $1/f$ noise leads to a Gaussian decay $e^{-(\Gamma_{\phi,1/f} t)^2}$. (b) Dephasing time vs modulation amplitude for various values of the white noise (additive and multiplicative) spectral densities: $A_{\white}= 10\,\mathrm{n}\Phi_0/\sqrt{\mathrm{Hz}}$ (teal), $50\,\mathrm{n}\Phi_0/\sqrt{\mathrm{Hz}}$ (blue), and $50\,\mathrm{n}\Phi_0/\sqrt{\mathrm{Hz}}$ with a lowpass filter (yellow). (c) Infidelity of a CZ02 gate vs flux modulation amplitude for fixed $1/f$ noise amplitude and for various values of white noise spectral densities mentioned in (a); the gray curve is the infidelity without decoherence. The flux pulse parameters are: $t_{\rm ramp} =10\units{ns}$, $t_f =\pi/g_{\rm eff}+2t_{\rm ramp}$ and we have assumed $T_{1,F} = T_{1,T}= 150\,\mu\mathrm{s},T_{2,F}^{*}=150~\mu \mathrm{s}$, and a background dephasing time for tunable qubit of $T_{\phi, \back}=300\,\mu\mathrm{s}$.}
	\label{fidelity_CZ20}
\end{figure}

We calculate the average process fidelity of a two-qubit gate using~\cite{Nielsen2002} 
\begin{align}
\mathcal{F} = \frac{\textrm{Tr}\{U^{\dag}_{\rm ideal}\mathcal{E}\}+d}{d(d+1)},
\end{align}
where $d$ is the dimension of the Hilbert space, $d=4$ for 2 qubits, $U_{\rm ideal}$ is the ideal process matrix for the target unitary gate, and $\mathcal{E}$ is the noisy process matrix. We optimize the fidelity by applying single-qubit rotations on each qubit, $R_{Z}(\theta_T)$ and $R_{Z}(\theta_F)$, optimizing the modulation frequency and the gate time. Note that in physical implementations the single qubit rotations depend on the details of the pulse (modulation frequency, amplitude, gate time, rise and fall times) and can be calibrated away.

To illustrate the contribution of dephasing on the performance of the parametrically activated CZ gate, we consider a tunable qubit of frequency at zero flux $\omega_T(0)/2\pi = 5.1\units{GHz}$, at half-flux quantum $\omega_T(\Phi_0/2)/2\pi = 4.5\units{GHz}$, and anharmonicity $\eta_{T}(0)/2\pi= 0.2\units{GHz}$ and fixed-frequency qubit of frequency $\omega_F/2\pi =4.0\units{GHz}$ and anharmonicity $\eta_F/2\pi = 0.2\units{GHz}$. The static coupling between the qubits is $g/2\pi = 7 \units{MHz}$. Let us assume that the tunable qubit is parked at zero flux, which is first-order flux insensitive to the $1/f$ DC flux noise. For a tunable qubit parked at $\Phi_\dc =0$, two CZ gates can be actuated by modulating the flux bias at $\wmod^{\rm CZ02} =|\wmean-\omega_{F}-\etamean|/2$  and $\wmod^{\rm CZ20}=|\wmean-\omega_{F}+\eta_F|/2$ and iSWAP at $\wmod^{\rm iSWAP}=|\wmean-\omega_{F}|/2$. The modulation at $\wmod^{\rm CZ02}$ creates a resonant interaction between $|11\rangle\leftrightarrow |02\rangle$ while the modulation at $\wmod^{\rm CZ20}$ enables a resonant interaction between $|11\rangle\leftrightarrow |20\rangle$. Here $\etamean$ is the average anharmonicity of the tunable qubit over a period of the modulation~\cite{Nico17}. 

The modulation frequencies for activating CZ and iSWAP gates are shown in Fig. \ref{modulation_frequency}(a). The dashed curves represent the second harmonics generated by the modulation. In a more connected qubits, these resonances might overlap with the fundamental harmonics activating the entangling gates. It is imperative to avoid collisions with higher order resonances to realize high fidelity entangling gates.  Note that the gate time at every operating point can be obtained from the effective coupling [Fig. \ref{modulation_frequency}(b)], which for CZ gates is given by
\begin{align}
t_{\rm CZ}(\Phiac) = \frac{\pi}{g_{\rm eff}(\Phiac)}+2t_{\rm ramp},
\end{align}
where for small modulation amplitude, the effective coupling is given by 
$g_{\rm eff}\approx\sqrt{2}gJ_{1}(\wF_2/2\wmod)$.

To determine the requirements of the white noise levels to achieve high-fidelity two-qubit gates at the \ACsweet{}, we computed the infidelity as function of the modulation amplitude for various levels of white noise and for a realistic $1/f$ noise amplitude. In Fig.~\ref{fidelity_CZ20}(a), we plot the total pure dephasing time due to $1/f$ flux noise vs flux modulation amplitude and Fig. \ref{fidelity_CZ20}(b) shows the dephasing time due to white noise. At the \ACsweet{}, $\Phiac\approx 0.6~\Phi_0$, the dephasing time is only limited by the white noise level and background dephasing. This is irrespective of the $1/f$ noise strength.

For a multiplicative $1/f$ flux noise of amplitude $A_{\ac,\pink}= 3.63~\mu\Phi_0$ and white noise (both additive and multiplicative) with spectral density of each $A_{\white} = 50\,\mathrm{n}\Phi_0/\sqrt{\mathrm{Hz}}$, the infidelity of the CZ02 gate is limited to $10^{-2}$ for $T_{1,F} = T_{1,T}= 150\,\mu\mathrm{s},T_{2,F}^{*}=150\,\mu\mathrm{s}$, and background dephasing time $T_{\phi,\back}=300~\mu\mathrm{s}$. When the strength of the white noise level is reduced to $A_{\white} = 10\,\mathrm{n}\Phi_0/\sqrt{\mathrm{Hz}}$, the infidelity improves to $\approx 10^{-3}$ which is limited by the incoherent errors. Instead of lowering the overall strength of the white noise, if we apply a lowpass filter (with cutoff $\sim 1.5\,\wpump$) on the flux bias line to filter out white noise coming from the electronics, we can recover the original \ACsweet{} at $\Phiac \approx 0.6~\Phi_0$. It is interesting to note that (for the same coherence times) the infidelity gradually decreases as the modulation amplitude is increased, reaching to $\approx 10^{-3}$ error level [see yellow curve in Fig.~\ref{fidelity_CZ20} (c)] at the \ACsweet{} obtained at a much weaker white noise level. 

Even in the presence of strong multiplicative $1/f$ flux noise and white noise, filtering the white noise by applying a lowpass filter  enables us to recover the \ACsweet{} and achieve high fidelity CZ gate. This is essentially the main result of the paper. 

\section{Conclusions}
\label{section_conclusion}

We have shown that tunable superconducting qubits under parametric modulation can have first-order flux insensitive point for low-frequency $1/f$ noise analogous to qubits without modulation. These flux sweet spots occur at the extrema of the qubit average frequency. Although the dephasing times at these flux sweet spots can be limited by the white noise, applying appropriate lowpass filter allows us to recover the sweet spots. This opens the door for realization of high fidelity parametrically activated entangling gates by operating these gates at the \ACsweet{} spot which otherwise are limited by the dephasing under modulation.

\section*{Contributions}

ND and EAS developed the theory and numerical modeling, JC developed the technique to perform numerical averaging 
of different types of flux noise. MPS supervised and coordinated the effort. All authors contributed to writing the manuscript.

\begin{acknowledgments}
We thank Sabrina Hong, Prasahnt Sivarajah, Colm Ryan, Alexander Papageorge, and Blake Johnson for helpful discussions. We also  thank Blake Johnson for critically reading the manuscript.
\end{acknowledgments}

\appendix

\section{Derivatives of Fourier series}
\label{appendix_Fourier}

Here it is more convenient to work in terms of phase $\phi$ ($2\pi$ periodic) rather than flux $\Phi$ ($\Phi_0$ periodic), the conversion is done through the flux quantum $\Phi_0=h/2e$: $\phi=2\pi\Phi/\Phi_0$.
The Fourier series of transmon frequency under modulation~\cite{Nico17} and their derivatives read,
\begin{align}
\wF_k&=(2-\delta_k)\sum_{n=0}^\infty s_n\cos(n\phidc+k\tfrac{\pi}{2})\BJ_k(n\phiac),\\
\frac{\partial\wF_k}{\partial\Phidc}&=(\delta_k-2)\sum_{n=0}^\infty ns_n\sin(n\phidc+k\tfrac{\pi}{2})\BJ_k(n\phiac),\label{nudc}\\
\frac{\partial\wF_k}{\partial\Phiac}&=(\delta_k-2)\sum_{n=0}^\infty ns_n\cos(n\phidc+k\tfrac{\pi}{2})\nonumber\\
&\qquad\qquad\quad\times\tfrac{1}{2}[\BJ_{k+1}(n\phiac)-\BJ_{k-1}(n\phiac)],\label{nuac}
\end{align}
with the Bessel functions $\BJ_k$.
The derivatives are plotted in Fig.~\ref{fig_Fourier} for illustration.
The parameter $s_n$ depends on the tunable transmon charging energy $E_C$ and Josephson energies $E_{J_1},E_{J_2}$,
\begin{multline}
s_n=\tfrac{1}{n!}(-\tfrac{1}{2}\mathcal{X})^n(2-\delta_n)E_C\sum_{p\in\mathbb{Z}}\omega^{(p)}\bar{\Xi}^\frac{p}{4}R_{n,p}\\
\times{}_2\mathrm{F}_1(\tfrac{n}{2}\!+\!\tfrac{p}{8},\tfrac{n+1}{2}\!+\!\tfrac{p}{8},n\!+\!1,\mathcal{X}^2),
\end{multline}
where ${}_2\mathrm{F}_1$ is the hypergeometric function
\begin{align}
\overline{\Xi}&=4E_C^2/(E_{J_1}^2+E_{J_2}^2), \\
\mathcal{X}&=2E_{J_1}E_{J_2}/(E_{J_1}^2+E_{J_2}^2), 
\end{align}
and
\[
  R_{n,p}=
  \begin{cases}
    0, & \textrm{if } p=0  ~\textrm{and}~ n>0\\
\frac{\Gamma(n+\frac{p}{4})}{\Gamma(\frac{p}{4})}&\mathrm{else}.
  \end{cases}
\]
Here $\Gamma$ is the gamma function. The terms $\omega^{(p)}$ are rational numbers derived from perturbation theory~\cite{Nico17}. The first 10 terms are:
\begin{align}
&\omega^{(-1)}=4,\quad
\omega^{(0)}=-1,\quad
\omega^{(1)}=-\frac{1}{2^2},\quad
\omega^{(2)}=-\frac{21}{2^7},\nonumber\\
&\omega^{(3)}=-\frac{19}{2^7},\quad
\omega^{(4)}=-\frac{5319}{2^{15}},\quad
\omega^{(5)}=-\frac{6649}{2^{15}},\nonumber\\
&\omega^{(6)}=-\frac{1180581}{2^{22}},\quad
\omega^{(7)}=-\frac{446287}{2^{20}},\nonumber\\
&\omega^{(8)}=-\frac{1489138635}{2^{31}}.
\end{align}

\section{Analytic derivation of dephasing rate under flux-bias modulation}
\label{appendix_rate}

In the frequency domain, the general expression of $\cor(t)$ is,
\begin{multline}
\cor(t)=\frac{1}{2}\sum_{k_1=0}^\infty\sum_{k_2=0}^\infty\int_{\wir}^{\wuv}\frac{\mathrm{d}\omega}{2\pi}I_{k_1,k_2}(\omega,t)\\
\times\left\{\dwFdc{k_1}\dwFdc{k_2}S_{\dc,\pink}(\omega)+\dwFac{k_1}\dwFac{k_2}S_{\ac,\pink}(\omega)\right\},
\end{multline}
with the time integral
\begin{multline}
I_{k_1,k_2}(\omega,t)=\int_0^t\mathrm{d}t_1\int_0^t\mathrm{d}t_2\cos[\omega(t_1-t_2)]\\\times\cos[k_1(\wmod t_1+\phimod)]\cos[k_2(\wmod t_2+\phimod)].
\label{timeintegral}
\end{multline}
The dephasing rate is extracted from the leading term of $\cor(t)$, it corresponds to the terms involving the time difference $t_1-t_2$ in Eq.~\eqref{timeintegral} that is then simplified to
\begin{align}
I_{k,k}(\omega,t)&=\tfrac{\delta_{k}}{4}t^2\left\{\sinc^2[\tfrac{1}{2}(\omega-k\wmod)t]+\sinc^2[\tfrac{1}{2}(\omega+k\wmod)t]\right\}\nonumber\\
&\stackrel{t\to\infty}{\to}\tfrac{\pi}{2}\delta_kt[\delta(\omega-k\wmod)+\delta(\omega+k\wmod)],
\end{align}
and yields Eq.~\eqref{simplecor}.
For $1/f$ noise, the parameter $\lambda$ in Eq.~\eqref{Gammapink} is calculated from \cite{Martinis2003},
\begin{align}
\int_{\wir}^{\wuv}\frac{\mathrm{d}\omega}{\omega}\sinc^2(\tfrac{1}{2}\omega t)\to\frac{3}{2}-\gamma-\ln(\wir t),
\end{align}
in the limit $\wir t\ll1$ and $\wuv t\gg1$.

White noise is characterized by a constant spectral density $A_\white$ and is, equivalently, delta-correlated in time domain,
$\langle \dPhi(t_1)\dPhi(t_2)\rangle=A_\white^2\delta(t_1-t_2)$.
This expression can be directly used in Eq.~\eqref{correlationdef} and yields,
\begin{align}
{\cor}_\white(t)&=\Gamma_{\phi,\white}t
+\sum_{k=1}^\infty B_k\{\sin[k(\wmod t+\theta_p)]-\sin(k\theta_p)\},
\end{align}
composed of a linear increase with the rate $\Gamma_{\phi,\white}$ of Eq.~\eqref{Gammawhite} and oscillations at harmonics of the modulation frequency with amplitude
\begin{align}
B_k
=\frac{A_{\dc,\white}}{4k\wmod}&\left(\sum_{l=0}^k\dwFdc{k-l}\dwFdc{l}+2\sum_{l=0}^\infty\dwFdc{k+l}\dwFdc{l}\right)\nonumber\\
+\frac{A_{\ac,\white}}{4k\wmod}&\left(\sum_{l=0}^k\dwFac{k-l}\dwFac{l}+2\sum_{l=0}^\infty\dwFac{k+l}\dwFac{l}\right).
\end{align}

\section{ Phenomenological time-dependent quantum master equation }
\label{appendix_QME}

Dephasing due to flux noise of a single three-level transmon leads to the following dynamics of the density matrix elements:
$\rho_{nm}(t)=\Exp{-\gamma_{\phi,nm}(t)}\rho_{nm}(0)$,
with $\gamma_{\phi,nm}(t)=(\Gamma_{\phi,nm}t)^\beta$
for $n,m\in\{0,1,2\}$ and 
$\Gamma_{\phi,nm} = \lambda A_{1/f}|\omega'_{nm}|^{2/\beta}$,
where the prime denotes derivation with respect to flux, $\omega'=\partial\omega/\partial\Phi$.
This temporal evolution of the density matrix is reproduced by the time-dependent Lindbladian 
$2\Gamma_{\phi,01}\dot{\gamma}_{\phi,01}(t)\mathcal{D}[|1\rangle\langle1|+2\Lambda|2\rangle\langle2|]$
with 
\begin{align}
\Lambda=\frac{1}{2}\left(\frac{\Gamma_{\phi,02}}{\Gamma_{\phi,01}}\right)^{\frac{\beta}{2}}
=\left|\frac{\omega_{02}'}{2\omega_{01}'}\right|
=\left|1-\frac{\eta'}{2\omega_{01}'}\right|
\approx1,
\end{align}
since, to leading order, 
$\eta'/\omega_{01}'\approx-(9/4)(\eta/\omega_{01})^2$ which is small in the transmon regime.
To describe the dephasing of the qutrit with one dissipator, we used the fact that $\omega'_{02}=\omega'_{01}+\omega'_{12}$, and hence $\Gamma_{\phi,02}^{\beta/2}=\Gamma_{\phi,01}^{\beta/2}+\Gamma_{\phi,12}^{\beta/2}$.
The time-dependent term in the quantum master equation Eq.~\eqref{tdQME} is then recovered.

\begin{figure}[t]
	\centering
	\includegraphics[width=\columnwidth]{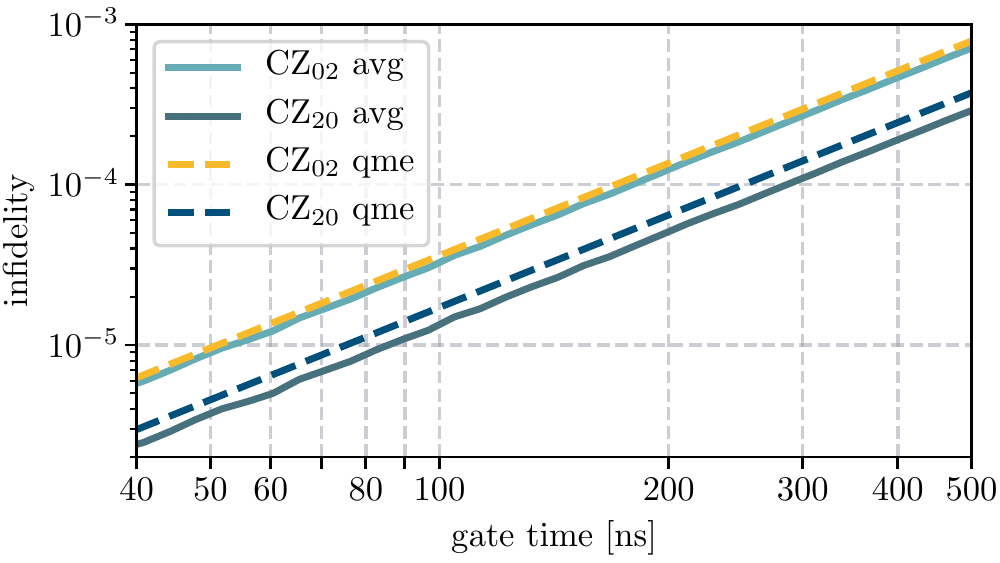}
	\caption{Average process fidelity for CZ obtained from averaging the coherent dynamics over flux noise (full lines) and obtained from the time-dependent quantum master equation (dashed lines). Parameters of the text.}
	\label{fig_qmevsavg}
\end{figure}

For an ideal Hamiltonian that couples the states $|11\rangle$ and $|02\rangle$ ($|20\rangle$) for CZ$_{02}$ (CZ$_{20}$) and considering only $1/f$ flux noise, the time-dependent quantum master equation can be solved in the limit where the effective coupling is much larger than the dephasing rate, the resulting average process fidelity reads,
\begin{align}
\mathcal{F}_{\mathrm{CZ}02}&\approx1-\frac{61}{80}\left(\frac{t_\mathrm{CZ}}{T_\phi}\right)^\beta,&
\mathcal{F}_{\mathrm{CZ}20}&\approx1-\frac{29}{80}\left(\frac{t_\mathrm{CZ}}{T_\phi}\right)^\beta,
\label{asymptoticfidelity}
\end{align}
for $t_\mathrm{CZ}\ll T_\phi$.
These values can be compared to the fidelity obtained from averaging the coherent dynamics of the entangling gate over flux noise.
The two approaches are in excellent agreement, the solution Eq.~\eqref{asymptoticfidelity} slightly underestimates the fidelity.
This is shown in Fig.~\ref{fig_qmevsavg}.
The average process fidelity is calculated for a tunable transmon with $\wT(\Phi=0)/2\pi =5.1\units{GHz}$, $\wT(\Phi=\Phi_0/2)/2\pi = 4.1\units{GHz}$,  $\eta_{T}(\Phi=0)/2\pi = 0.2\units{GHz}$ which is subjected to $1/f$ flux noise characterized by a dephasing rate $T_\phi=18\,\mu\mathrm{s}$ and a coefficient $\beta=1.9$.
We consider the ideal Hamiltonian in the interaction picture,
$H_{\mathrm{CZ}02}=\delta\omega_{T}(t) [|01\rangle\langle 01|+|11\rangle\langle 11|] +[2\delta\omega_{T}(t)-\delta\eta_T(t)] |02\rangle \langle02| +g_\eff[|11\rangle\langle02]+|02\rangle\langle11|]$ (and similarly for CZ$_{20}$).
The effective coupling is varied from $1\,\mathrm{MHz}$ to $12.5\,\mathrm{MHz}$ to artificially change the gate time between $t_\mathrm{CZ}=40\,\mathrm{ns}$ and $t_\mathrm{CZ}=500\,\mathrm{ns}$. The modulation frequency is set to $\omega_m/(2\pi)=300\,\mathrm{MHz}$.
The fidelity obtained from the averaged coherent dynamics and the asymptotic fidelity Eq.~\eqref{asymptoticfidelity} are plotted as a function of $t_\mathrm{CZ}/T_\phi$.

\begin{thebibliography}{35}%
\makeatletter
\providecommand \@ifxundefined [1]{%
 \@ifx{#1\undefined}
}%
\providecommand \@ifnum [1]{%
 \ifnum #1\expandafter \@firstoftwo
 \else \expandafter \@secondoftwo
 \fi
}%
\providecommand \@ifx [1]{%
 \ifx #1\expandafter \@firstoftwo
 \else \expandafter \@secondoftwo
 \fi
}%
\providecommand \natexlab [1]{#1}%
\providecommand \enquote  [1]{``#1''}%
\providecommand \bibnamefont  [1]{#1}%
\providecommand \bibfnamefont [1]{#1}%
\providecommand \citenamefont [1]{#1}%
\providecommand \href@noop [0]{\@secondoftwo}%
\providecommand \href [0]{\begingroup \@sanitize@url \@href}%
\providecommand \@href[1]{\@@startlink{#1}\@@href}%
\providecommand \@@href[1]{\endgroup#1\@@endlink}%
\providecommand \@sanitize@url [0]{\catcode `\\12\catcode `\$12\catcode
  `\&12\catcode `\#12\catcode `\^12\catcode `\_12\catcode `\%12\relax}%
\providecommand \@@startlink[1]{}%
\providecommand \@@endlink[0]{}%
\providecommand \url  [0]{\begingroup\@sanitize@url \@url }%
\providecommand \@url [1]{\endgroup\@href {#1}{\urlprefix }}%
\providecommand \urlprefix  [0]{URL }%
\providecommand \Eprint [0]{\href }%
\providecommand \doibase [0]{http://dx.doi.org/}%
\providecommand \selectlanguage [0]{\@gobble}%
\providecommand \bibinfo  [0]{\@secondoftwo}%
\providecommand \bibfield  [0]{\@secondoftwo}%
\providecommand \translation [1]{[#1]}%
\providecommand \BibitemOpen [0]{}%
\providecommand \bibitemStop [0]{}%
\providecommand \bibitemNoStop [0]{.\EOS\space}%
\providecommand \EOS [0]{\spacefactor3000\relax}%
\providecommand \BibitemShut  [1]{\csname bibitem#1\endcsname}%
\let\auto@bib@innerbib\@empty
\bibitem [{\citenamefont {Koch}\ \emph {et~al.}(1983)\citenamefont {Koch},
  \citenamefont {Clarke}, \citenamefont {Goubau}, \citenamefont {Martinis},
  \citenamefont {Pegrum},\ and\ \citenamefont {van Harlingen}}]{Koch1983}%
  \BibitemOpen
  \bibfield  {author} {\bibinfo {author} {\bibfnamefont {R.~H.}\ \bibnamefont
  {Koch}}, \bibinfo {author} {\bibfnamefont {J.}~\bibnamefont {Clarke}},
  \bibinfo {author} {\bibfnamefont {W.~M.}\ \bibnamefont {Goubau}}, \bibinfo
  {author} {\bibfnamefont {J.~M.}\ \bibnamefont {Martinis}}, \bibinfo {author}
  {\bibfnamefont {C.~M.}\ \bibnamefont {Pegrum}}, \ and\ \bibinfo {author}
  {\bibfnamefont {D.~J.}\ \bibnamefont {van Harlingen}},\ }\href@noop {}
  {\bibfield  {journal} {\bibinfo  {journal} {Journal of Low Temperature
  Physics}\ }\textbf {\bibinfo {volume} {51}},\ \bibinfo {pages} {207}
  (\bibinfo {year} {1983})}\BibitemShut {NoStop}%
\bibitem [{\citenamefont {Wellstood}\ \emph {et~al.}(1987)\citenamefont
  {Wellstood}, \citenamefont {Urbina},\ and\ \citenamefont
  {Clarke}}]{Wellstood87}%
  \BibitemOpen
  \bibfield  {author} {\bibinfo {author} {\bibfnamefont {F.~C.}\ \bibnamefont
  {Wellstood}}, \bibinfo {author} {\bibfnamefont {C.}~\bibnamefont {Urbina}}, \
  and\ \bibinfo {author} {\bibfnamefont {J.}~\bibnamefont {Clarke}},\
  }\href@noop {} {\bibfield  {journal} {\bibinfo  {journal} {Applied Physics
  Letters}\ }\textbf {\bibinfo {volume} {50}},\ \bibinfo {pages} {772}
  (\bibinfo {year} {1987})}\BibitemShut {NoStop}%
\bibitem [{\citenamefont {Vion}\ \emph {et~al.}(2002)\citenamefont {Vion},
  \citenamefont {Aassime}, \citenamefont {Cottet}, \citenamefont {Joyez},
  \citenamefont {Pothier}, \citenamefont {Urbina}, \citenamefont {Esteve},\
  and\ \citenamefont {Devoret}}]{Devoret2002}%
  \BibitemOpen
  \bibfield  {author} {\bibinfo {author} {\bibfnamefont {D.}~\bibnamefont
  {Vion}}, \bibinfo {author} {\bibfnamefont {A.}~\bibnamefont {Aassime}},
  \bibinfo {author} {\bibfnamefont {A.}~\bibnamefont {Cottet}}, \bibinfo
  {author} {\bibfnamefont {P.}~\bibnamefont {Joyez}}, \bibinfo {author}
  {\bibfnamefont {H.}~\bibnamefont {Pothier}}, \bibinfo {author} {\bibfnamefont
  {C.}~\bibnamefont {Urbina}}, \bibinfo {author} {\bibfnamefont
  {D.}~\bibnamefont {Esteve}}, \ and\ \bibinfo {author} {\bibfnamefont {M.~H.}\
  \bibnamefont {Devoret}},\ }\href {\doibase 10.1126/science.1069372}
  {\bibfield  {journal} {\bibinfo  {journal} {Science}\ }\textbf {\bibinfo
  {volume} {296}},\ \bibinfo {pages} {886} (\bibinfo {year}
  {2002})}\BibitemShut {NoStop}%
\bibitem [{\citenamefont {Martinis}\ \emph {et~al.}(2003)\citenamefont
  {Martinis}, \citenamefont {Nam}, \citenamefont {Aumentado}, \citenamefont
  {Lang},\ and\ \citenamefont {Urbina}}]{Martinis2003}%
  \BibitemOpen
  \bibfield  {author} {\bibinfo {author} {\bibfnamefont {J.~M.}\ \bibnamefont
  {Martinis}}, \bibinfo {author} {\bibfnamefont {S.}~\bibnamefont {Nam}},
  \bibinfo {author} {\bibfnamefont {J.}~\bibnamefont {Aumentado}}, \bibinfo
  {author} {\bibfnamefont {K.~M.}\ \bibnamefont {Lang}}, \ and\ \bibinfo
  {author} {\bibfnamefont {C.}~\bibnamefont {Urbina}},\ }\href {\doibase
  10.1103/PhysRevB.67.094510} {\bibfield  {journal} {\bibinfo  {journal} {Phys.
  Rev. B}\ }\textbf {\bibinfo {volume} {67}},\ \bibinfo {pages} {094510}
  (\bibinfo {year} {2003})}\BibitemShut {NoStop}%
\bibitem [{\citenamefont {Ithier}\ \emph {et~al.}(2005)\citenamefont {Ithier},
  \citenamefont {Collin}, \citenamefont {Joyez}, \citenamefont {Meeson},
  \citenamefont {Vion}, \citenamefont {Esteve}, \citenamefont {Chiarello},
  \citenamefont {Shnirman}, \citenamefont {Makhlin}, \citenamefont {Schriefl},\
  and\ \citenamefont {Sch\"on}}]{Ithier05}%
  \BibitemOpen
  \bibfield  {author} {\bibinfo {author} {\bibfnamefont {G.}~\bibnamefont
  {Ithier}}, \bibinfo {author} {\bibfnamefont {E.}~\bibnamefont {Collin}},
  \bibinfo {author} {\bibfnamefont {P.}~\bibnamefont {Joyez}}, \bibinfo
  {author} {\bibfnamefont {P.~J.}\ \bibnamefont {Meeson}}, \bibinfo {author}
  {\bibfnamefont {D.}~\bibnamefont {Vion}}, \bibinfo {author} {\bibfnamefont
  {D.}~\bibnamefont {Esteve}}, \bibinfo {author} {\bibfnamefont
  {F.}~\bibnamefont {Chiarello}}, \bibinfo {author} {\bibfnamefont
  {A.}~\bibnamefont {Shnirman}}, \bibinfo {author} {\bibfnamefont
  {Y.}~\bibnamefont {Makhlin}}, \bibinfo {author} {\bibfnamefont
  {J.}~\bibnamefont {Schriefl}}, \ and\ \bibinfo {author} {\bibfnamefont
  {G.}~\bibnamefont {Sch\"on}},\ }\href@noop {} {\bibfield  {journal} {\bibinfo
   {journal} {Phys. Rev. B}\ }\textbf {\bibinfo {volume} {72}},\ \bibinfo
  {pages} {134519} (\bibinfo {year} {2005})}\BibitemShut {NoStop}%
\bibitem [{\citenamefont {Yoshihara}\ \emph {et~al.}(2006)\citenamefont
  {Yoshihara}, \citenamefont {Harrabi}, \citenamefont {Niskanen}, \citenamefont
  {Nakamura},\ and\ \citenamefont {Tsai}}]{Yoshihara_2006}%
  \BibitemOpen
  \bibfield  {author} {\bibinfo {author} {\bibfnamefont {F.}~\bibnamefont
  {Yoshihara}}, \bibinfo {author} {\bibfnamefont {K.}~\bibnamefont {Harrabi}},
  \bibinfo {author} {\bibfnamefont {A.~O.}\ \bibnamefont {Niskanen}}, \bibinfo
  {author} {\bibfnamefont {Y.}~\bibnamefont {Nakamura}}, \ and\ \bibinfo
  {author} {\bibfnamefont {J.~S.}\ \bibnamefont {Tsai}},\ }\href {\doibase
  10.1103/PhysRevLett.97.167001} {\bibfield  {journal} {\bibinfo  {journal}
  {Phys. Rev. Lett.}\ }\textbf {\bibinfo {volume} {97}},\ \bibinfo {pages}
  {167001} (\bibinfo {year} {2006})}\BibitemShut {NoStop}%
\bibitem [{\citenamefont {Bialczak}\ \emph {et~al.}(2007)\citenamefont
  {Bialczak}, \citenamefont {McDermott}, \citenamefont {Ansmann}, \citenamefont
  {Hofheinz}, \citenamefont {Katz}, \citenamefont {Lucero}, \citenamefont
  {Neeley}, \citenamefont {O'Connell}, \citenamefont {Wang}, \citenamefont
  {Cleland},\ and\ \citenamefont {Martinis}}]{Bialczak_2007}%
  \BibitemOpen
  \bibfield  {author} {\bibinfo {author} {\bibfnamefont {R.~C.}\ \bibnamefont
  {Bialczak}}, \bibinfo {author} {\bibfnamefont {R.}~\bibnamefont {McDermott}},
  \bibinfo {author} {\bibfnamefont {M.}~\bibnamefont {Ansmann}}, \bibinfo
  {author} {\bibfnamefont {M.}~\bibnamefont {Hofheinz}}, \bibinfo {author}
  {\bibfnamefont {N.}~\bibnamefont {Katz}}, \bibinfo {author} {\bibfnamefont
  {E.}~\bibnamefont {Lucero}}, \bibinfo {author} {\bibfnamefont
  {M.}~\bibnamefont {Neeley}}, \bibinfo {author} {\bibfnamefont {A.~D.}\
  \bibnamefont {O'Connell}}, \bibinfo {author} {\bibfnamefont {H.}~\bibnamefont
  {Wang}}, \bibinfo {author} {\bibfnamefont {A.~N.}\ \bibnamefont {Cleland}}, \
  and\ \bibinfo {author} {\bibfnamefont {J.~M.}\ \bibnamefont {Martinis}},\
  }\href {\doibase 10.1103/PhysRevLett.99.187006} {\bibfield  {journal}
  {\bibinfo  {journal} {Phys. Rev. Lett.}\ }\textbf {\bibinfo {volume} {99}},\
  \bibinfo {pages} {187006} (\bibinfo {year} {2007})}\BibitemShut {NoStop}%
\bibitem [{\citenamefont {Koch}\ \emph {et~al.}(2007)\citenamefont {Koch},
  \citenamefont {DiVincenzo},\ and\ \citenamefont {Clarke}}]{RHKoch_2007}%
  \BibitemOpen
  \bibfield  {author} {\bibinfo {author} {\bibfnamefont {R.~H.}\ \bibnamefont
  {Koch}}, \bibinfo {author} {\bibfnamefont {D.~P.}\ \bibnamefont
  {DiVincenzo}}, \ and\ \bibinfo {author} {\bibfnamefont {J.}~\bibnamefont
  {Clarke}},\ }\href {\doibase 10.1103/PhysRevLett.98.267003} {\bibfield
  {journal} {\bibinfo  {journal} {Phys. Rev. Lett.}\ }\textbf {\bibinfo
  {volume} {98}},\ \bibinfo {pages} {267003} (\bibinfo {year}
  {2007})}\BibitemShut {NoStop}%
\bibitem [{\citenamefont {Faoro}\ and\ \citenamefont
  {Ioffe}(2008)}]{Faoro_2008}%
  \BibitemOpen
  \bibfield  {author} {\bibinfo {author} {\bibfnamefont {L.}~\bibnamefont
  {Faoro}}\ and\ \bibinfo {author} {\bibfnamefont {L.~B.}\ \bibnamefont
  {Ioffe}},\ }\href {\doibase 10.1103/PhysRevLett.100.227005} {\bibfield
  {journal} {\bibinfo  {journal} {Phys. Rev. Lett.}\ }\textbf {\bibinfo
  {volume} {100}},\ \bibinfo {pages} {227005} (\bibinfo {year}
  {2008})}\BibitemShut {NoStop}%
\bibitem [{\citenamefont {Manucharyan}\ \emph {et~al.}(2009)\citenamefont
  {Manucharyan}, \citenamefont {Koch}, \citenamefont {Glazman},\ and\
  \citenamefont {Devoret}}]{Manucharyan_2009}%
  \BibitemOpen
  \bibfield  {author} {\bibinfo {author} {\bibfnamefont {V.~E.}\ \bibnamefont
  {Manucharyan}}, \bibinfo {author} {\bibfnamefont {J.}~\bibnamefont {Koch}},
  \bibinfo {author} {\bibfnamefont {L.~I.}\ \bibnamefont {Glazman}}, \ and\
  \bibinfo {author} {\bibfnamefont {M.~H.}\ \bibnamefont {Devoret}},\ }\href
  {http://science.sciencemag.org/content/326/5949/113.abstract} {\bibfield
  {journal} {\bibinfo  {journal} {Science}\ }\textbf {\bibinfo {volume}
  {326}},\ \bibinfo {pages} {113} (\bibinfo {year} {2009})}\BibitemShut
  {NoStop}%
\bibitem [{\citenamefont {Barends}\ \emph {et~al.}(2013)\citenamefont
  {Barends}, \citenamefont {Kelly}, \citenamefont {Megrant}, \citenamefont
  {Sank}, \citenamefont {Jeffrey}, \citenamefont {Chen}, \citenamefont {Yin},
  \citenamefont {Chiaro}, \citenamefont {Mutus}, \citenamefont {Neill},
  \citenamefont {O'Malley}, \citenamefont {Roushan}, \citenamefont {Wenner},
  \citenamefont {White}, \citenamefont {Cleland},\ and\ \citenamefont
  {Martinis}}]{xmon}%
  \BibitemOpen
  \bibfield  {author} {\bibinfo {author} {\bibfnamefont {R.}~\bibnamefont
  {Barends}}, \bibinfo {author} {\bibfnamefont {J.}~\bibnamefont {Kelly}},
  \bibinfo {author} {\bibfnamefont {A.}~\bibnamefont {Megrant}}, \bibinfo
  {author} {\bibfnamefont {D.}~\bibnamefont {Sank}}, \bibinfo {author}
  {\bibfnamefont {E.}~\bibnamefont {Jeffrey}}, \bibinfo {author} {\bibfnamefont
  {Y.}~\bibnamefont {Chen}}, \bibinfo {author} {\bibfnamefont {Y.}~\bibnamefont
  {Yin}}, \bibinfo {author} {\bibfnamefont {B.}~\bibnamefont {Chiaro}},
  \bibinfo {author} {\bibfnamefont {J.}~\bibnamefont {Mutus}}, \bibinfo
  {author} {\bibfnamefont {C.}~\bibnamefont {Neill}}, \bibinfo {author}
  {\bibfnamefont {P.}~\bibnamefont {O'Malley}}, \bibinfo {author}
  {\bibfnamefont {P.}~\bibnamefont {Roushan}}, \bibinfo {author} {\bibfnamefont
  {J.}~\bibnamefont {Wenner}}, \bibinfo {author} {\bibfnamefont {T.~C.}\
  \bibnamefont {White}}, \bibinfo {author} {\bibfnamefont {A.~N.}\ \bibnamefont
  {Cleland}}, \ and\ \bibinfo {author} {\bibfnamefont {J.~M.}\ \bibnamefont
  {Martinis}},\ }\href {\doibase 10.1103/PhysRevLett.111.080502} {\bibfield
  {journal} {\bibinfo  {journal} {Phys. Rev. Lett.}\ }\textbf {\bibinfo
  {volume} {111}},\ \bibinfo {pages} {080502} (\bibinfo {year}
  {2013})}\BibitemShut {NoStop}%
\bibitem [{\citenamefont {Martinis}\ and\ \citenamefont
  {Megrant}()}]{Martinis_2014}%
  \BibitemOpen
  \bibfield  {author} {\bibinfo {author} {\bibfnamefont {J.~M.}\ \bibnamefont
  {Martinis}}\ and\ \bibinfo {author} {\bibfnamefont {A.}~\bibnamefont
  {Megrant}},\ }\href@noop {} {\enquote {\bibinfo {title} {Ucsb final report
  for the csq program: Review of decoherence and materials physics for
  superconducting qubits},}\ }\Eprint {http://arxiv.org/abs/arXiv:1410.5793}
  {arXiv:1410.5793} \BibitemShut {NoStop}%
\bibitem [{\citenamefont {Wang}\ \emph {et~al.}(2015)\citenamefont {Wang},
  \citenamefont {Shi}, \citenamefont {Hu}, \citenamefont {Han}, \citenamefont
  {Yu},\ and\ \citenamefont {Wu}}]{Wang15}%
  \BibitemOpen
  \bibfield  {author} {\bibinfo {author} {\bibfnamefont {H.}~\bibnamefont
  {Wang}}, \bibinfo {author} {\bibfnamefont {C.}~\bibnamefont {Shi}}, \bibinfo
  {author} {\bibfnamefont {J.}~\bibnamefont {Hu}}, \bibinfo {author}
  {\bibfnamefont {S.}~\bibnamefont {Han}}, \bibinfo {author} {\bibfnamefont
  {C.~C.}\ \bibnamefont {Yu}}, \ and\ \bibinfo {author} {\bibfnamefont {R.~Q.}\
  \bibnamefont {Wu}},\ }\href {\doibase 10.1103/PhysRevLett.115.077002}
  {\bibfield  {journal} {\bibinfo  {journal} {Phys. Rev. Lett.}\ }\textbf
  {\bibinfo {volume} {115}},\ \bibinfo {pages} {077002} (\bibinfo {year}
  {2015})}\BibitemShut {NoStop}%
\bibitem [{\citenamefont {O'Malley}\ \emph {et~al.}(2015)\citenamefont
  {O'Malley}, \citenamefont {Kelly}, \citenamefont {Barends}, \citenamefont
  {Campbell}, \citenamefont {Chen}, \citenamefont {Chen}, \citenamefont
  {Chiaro}, \citenamefont {Dunsworth}, \citenamefont {Fowler}, \citenamefont
  {Hoi}, \citenamefont {Jeffrey}, \citenamefont {Megrant}, \citenamefont
  {Mutus}, \citenamefont {Neill}, \citenamefont {Quintana}, \citenamefont
  {Roushan}, \citenamefont {Sank}, \citenamefont {Vainsencher}, \citenamefont
  {Wenner}, \citenamefont {White}, \citenamefont {Korotkov}, \citenamefont
  {Cleland},\ and\ \citenamefont {Martinis}}]{Omalley15}%
  \BibitemOpen
  \bibfield  {author} {\bibinfo {author} {\bibfnamefont {P.~J.~J.}\
  \bibnamefont {O'Malley}}, \bibinfo {author} {\bibfnamefont {J.}~\bibnamefont
  {Kelly}}, \bibinfo {author} {\bibfnamefont {R.}~\bibnamefont {Barends}},
  \bibinfo {author} {\bibfnamefont {B.}~\bibnamefont {Campbell}}, \bibinfo
  {author} {\bibfnamefont {Y.}~\bibnamefont {Chen}}, \bibinfo {author}
  {\bibfnamefont {Z.}~\bibnamefont {Chen}}, \bibinfo {author} {\bibfnamefont
  {B.}~\bibnamefont {Chiaro}}, \bibinfo {author} {\bibfnamefont
  {A.}~\bibnamefont {Dunsworth}}, \bibinfo {author} {\bibfnamefont {A.~G.}\
  \bibnamefont {Fowler}}, \bibinfo {author} {\bibfnamefont {I.-C.}\
  \bibnamefont {Hoi}}, \bibinfo {author} {\bibfnamefont {E.}~\bibnamefont
  {Jeffrey}}, \bibinfo {author} {\bibfnamefont {A.}~\bibnamefont {Megrant}},
  \bibinfo {author} {\bibfnamefont {J.}~\bibnamefont {Mutus}}, \bibinfo
  {author} {\bibfnamefont {C.}~\bibnamefont {Neill}}, \bibinfo {author}
  {\bibfnamefont {C.}~\bibnamefont {Quintana}}, \bibinfo {author}
  {\bibfnamefont {P.}~\bibnamefont {Roushan}}, \bibinfo {author} {\bibfnamefont
  {D.}~\bibnamefont {Sank}}, \bibinfo {author} {\bibfnamefont {A.}~\bibnamefont
  {Vainsencher}}, \bibinfo {author} {\bibfnamefont {J.}~\bibnamefont {Wenner}},
  \bibinfo {author} {\bibfnamefont {T.~C.}\ \bibnamefont {White}}, \bibinfo
  {author} {\bibfnamefont {A.~N.}\ \bibnamefont {Korotkov}}, \bibinfo {author}
  {\bibfnamefont {A.~N.}\ \bibnamefont {Cleland}}, \ and\ \bibinfo {author}
  {\bibfnamefont {J.~M.}\ \bibnamefont {Martinis}},\ }\href {\doibase
  10.1103/PhysRevApplied.3.044009} {\bibfield  {journal} {\bibinfo  {journal}
  {Phys. Rev. Applied}\ }\textbf {\bibinfo {volume} {3}},\ \bibinfo {pages}
  {044009} (\bibinfo {year} {2015})}\BibitemShut {NoStop}%
\bibitem [{\citenamefont {Kumar}\ \emph {et~al.}(2016)\citenamefont {Kumar},
  \citenamefont {Sendelbach}, \citenamefont {Beck}, \citenamefont {Freeland},
  \citenamefont {Wang}, \citenamefont {Wang}, \citenamefont {Yu}, \citenamefont
  {Wu}, \citenamefont {Pappas},\ and\ \citenamefont {McDermott}}]{Kumar16}%
  \BibitemOpen
  \bibfield  {author} {\bibinfo {author} {\bibfnamefont {P.}~\bibnamefont
  {Kumar}}, \bibinfo {author} {\bibfnamefont {S.}~\bibnamefont {Sendelbach}},
  \bibinfo {author} {\bibfnamefont {M.~A.}\ \bibnamefont {Beck}}, \bibinfo
  {author} {\bibfnamefont {J.~W.}\ \bibnamefont {Freeland}}, \bibinfo {author}
  {\bibfnamefont {Z.}~\bibnamefont {Wang}}, \bibinfo {author} {\bibfnamefont
  {H.}~\bibnamefont {Wang}}, \bibinfo {author} {\bibfnamefont {C.~C.}\
  \bibnamefont {Yu}}, \bibinfo {author} {\bibfnamefont {R.~Q.}\ \bibnamefont
  {Wu}}, \bibinfo {author} {\bibfnamefont {D.~P.}\ \bibnamefont {Pappas}}, \
  and\ \bibinfo {author} {\bibfnamefont {R.}~\bibnamefont {McDermott}},\
  }\href@noop {} {\bibfield  {journal} {\bibinfo  {journal} {Phys. Rev.
  Applied}\ }\textbf {\bibinfo {volume} {6}},\ \bibinfo {pages} {041001}
  (\bibinfo {year} {2016})}\BibitemShut {NoStop}%
\bibitem [{\citenamefont {Yan}\ \emph {et~al.}(2016)\citenamefont {Yan},
  \citenamefont {Gustavsson}, \citenamefont {Kamal}, \citenamefont {Birenbaum},
  \citenamefont {Sears}, \citenamefont {Hover}, \citenamefont {Gudmundsen},
  \citenamefont {Rosenberg}, \citenamefont {Samach}, \citenamefont {Weber},
  \citenamefont {Yoder}, \citenamefont {Orlando}, \citenamefont {Clarke},
  \citenamefont {Kerman},\ and\ \citenamefont {Oliver}}]{Yan_2016}%
  \BibitemOpen
  \bibfield  {author} {\bibinfo {author} {\bibfnamefont {F.}~\bibnamefont
  {Yan}}, \bibinfo {author} {\bibfnamefont {S.}~\bibnamefont {Gustavsson}},
  \bibinfo {author} {\bibfnamefont {A.}~\bibnamefont {Kamal}}, \bibinfo
  {author} {\bibfnamefont {J.}~\bibnamefont {Birenbaum}}, \bibinfo {author}
  {\bibfnamefont {A.~P.}\ \bibnamefont {Sears}}, \bibinfo {author}
  {\bibfnamefont {D.}~\bibnamefont {Hover}}, \bibinfo {author} {\bibfnamefont
  {T.~J.}\ \bibnamefont {Gudmundsen}}, \bibinfo {author} {\bibfnamefont
  {D.}~\bibnamefont {Rosenberg}}, \bibinfo {author} {\bibfnamefont
  {G.}~\bibnamefont {Samach}}, \bibinfo {author} {\bibfnamefont
  {S.}~\bibnamefont {Weber}}, \bibinfo {author} {\bibfnamefont {J.~L.}\
  \bibnamefont {Yoder}}, \bibinfo {author} {\bibfnamefont {T.~P.}\ \bibnamefont
  {Orlando}}, \bibinfo {author} {\bibfnamefont {J.}~\bibnamefont {Clarke}},
  \bibinfo {author} {\bibfnamefont {A.~J.}\ \bibnamefont {Kerman}}, \ and\
  \bibinfo {author} {\bibfnamefont {W.~D.}\ \bibnamefont {Oliver}},\
  }\href@noop {} {\bibfield  {journal} {\bibinfo  {journal} {Nature
  Communications}\ }\textbf {\bibinfo {volume} {7}},\ \bibinfo {pages} {12964}
  (\bibinfo {year} {2016})}\BibitemShut {NoStop}%
\bibitem [{\citenamefont {Quintana}\ \emph {et~al.}(2017)\citenamefont
  {Quintana}, \citenamefont {Chen}, \citenamefont {Sank}, \citenamefont
  {Petukhov}, \citenamefont {White}, \citenamefont {Kafri}, \citenamefont
  {Chiaro}, \citenamefont {Megrant}, \citenamefont {Barends}, \citenamefont
  {Campbell}, \citenamefont {Chen}, \citenamefont {Dunsworth}, \citenamefont
  {Fowler}, \citenamefont {Graff}, \citenamefont {Jeffrey}, \citenamefont
  {Kelly}, \citenamefont {Lucero}, \citenamefont {Mutus}, \citenamefont
  {Neeley}, \citenamefont {Neill}, \citenamefont {O'Malley}, \citenamefont
  {Roushan}, \citenamefont {Shabani}, \citenamefont {Smelyanskiy},
  \citenamefont {Vainsencher}, \citenamefont {Wenner}, \citenamefont {Neven},\
  and\ \citenamefont {Martinis}}]{Quintana_2017}%
  \BibitemOpen
  \bibfield  {author} {\bibinfo {author} {\bibfnamefont {C.~M.}\ \bibnamefont
  {Quintana}}, \bibinfo {author} {\bibfnamefont {Y.}~\bibnamefont {Chen}},
  \bibinfo {author} {\bibfnamefont {D.}~\bibnamefont {Sank}}, \bibinfo {author}
  {\bibfnamefont {A.~G.}\ \bibnamefont {Petukhov}}, \bibinfo {author}
  {\bibfnamefont {T.~C.}\ \bibnamefont {White}}, \bibinfo {author}
  {\bibfnamefont {D.}~\bibnamefont {Kafri}}, \bibinfo {author} {\bibfnamefont
  {B.}~\bibnamefont {Chiaro}}, \bibinfo {author} {\bibfnamefont
  {A.}~\bibnamefont {Megrant}}, \bibinfo {author} {\bibfnamefont
  {R.}~\bibnamefont {Barends}}, \bibinfo {author} {\bibfnamefont
  {B.}~\bibnamefont {Campbell}}, \bibinfo {author} {\bibfnamefont
  {Z.}~\bibnamefont {Chen}}, \bibinfo {author} {\bibfnamefont {A.}~\bibnamefont
  {Dunsworth}}, \bibinfo {author} {\bibfnamefont {A.~G.}\ \bibnamefont
  {Fowler}}, \bibinfo {author} {\bibfnamefont {R.}~\bibnamefont {Graff}},
  \bibinfo {author} {\bibfnamefont {E.}~\bibnamefont {Jeffrey}}, \bibinfo
  {author} {\bibfnamefont {J.}~\bibnamefont {Kelly}}, \bibinfo {author}
  {\bibfnamefont {E.}~\bibnamefont {Lucero}}, \bibinfo {author} {\bibfnamefont
  {J.~Y.}\ \bibnamefont {Mutus}}, \bibinfo {author} {\bibfnamefont
  {M.}~\bibnamefont {Neeley}}, \bibinfo {author} {\bibfnamefont
  {C.}~\bibnamefont {Neill}}, \bibinfo {author} {\bibfnamefont {P.~J.~J.}\
  \bibnamefont {O'Malley}}, \bibinfo {author} {\bibfnamefont {P.}~\bibnamefont
  {Roushan}}, \bibinfo {author} {\bibfnamefont {A.}~\bibnamefont {Shabani}},
  \bibinfo {author} {\bibfnamefont {V.~N.}\ \bibnamefont {Smelyanskiy}},
  \bibinfo {author} {\bibfnamefont {A.}~\bibnamefont {Vainsencher}}, \bibinfo
  {author} {\bibfnamefont {J.}~\bibnamefont {Wenner}}, \bibinfo {author}
  {\bibfnamefont {H.}~\bibnamefont {Neven}}, \ and\ \bibinfo {author}
  {\bibfnamefont {J.~M.}\ \bibnamefont {Martinis}},\ }\href {\doibase
  10.1103/PhysRevLett.118.057702} {\bibfield  {journal} {\bibinfo  {journal}
  {Phys. Rev. Lett.}\ }\textbf {\bibinfo {volume} {118}},\ \bibinfo {pages}
  {057702} (\bibinfo {year} {2017})}\BibitemShut {NoStop}%
\bibitem [{\citenamefont {Kou}\ \emph {et~al.}(2017)\citenamefont {Kou},
  \citenamefont {Smith}, \citenamefont {Vool}, \citenamefont {Brierley},
  \citenamefont {Meier}, \citenamefont {Frunzio}, \citenamefont {Girvin},
  \citenamefont {Glazman},\ and\ \citenamefont {Devoret}}]{Kou_2017}%
  \BibitemOpen
  \bibfield  {author} {\bibinfo {author} {\bibfnamefont {A.}~\bibnamefont
  {Kou}}, \bibinfo {author} {\bibfnamefont {W.~C.}\ \bibnamefont {Smith}},
  \bibinfo {author} {\bibfnamefont {U.}~\bibnamefont {Vool}}, \bibinfo {author}
  {\bibfnamefont {R.~T.}\ \bibnamefont {Brierley}}, \bibinfo {author}
  {\bibfnamefont {H.}~\bibnamefont {Meier}}, \bibinfo {author} {\bibfnamefont
  {L.}~\bibnamefont {Frunzio}}, \bibinfo {author} {\bibfnamefont {S.~M.}\
  \bibnamefont {Girvin}}, \bibinfo {author} {\bibfnamefont {L.~I.}\
  \bibnamefont {Glazman}}, \ and\ \bibinfo {author} {\bibfnamefont {M.~H.}\
  \bibnamefont {Devoret}},\ }\href {\doibase 10.1103/PhysRevX.7.031037}
  {\bibfield  {journal} {\bibinfo  {journal} {Phys. Rev. X}\ }\textbf {\bibinfo
  {volume} {7}},\ \bibinfo {pages} {031037} (\bibinfo {year}
  {2017})}\BibitemShut {NoStop}%
\bibitem [{\citenamefont {Hutchings}\ \emph {et~al.}(2017)\citenamefont
  {Hutchings}, \citenamefont {Hertzberg}, \citenamefont {Liu}, \citenamefont
  {Bronn}, \citenamefont {Keefe}, \citenamefont {Brink}, \citenamefont {Chow},\
  and\ \citenamefont {Plourde}}]{Plourde17}%
  \BibitemOpen
  \bibfield  {author} {\bibinfo {author} {\bibfnamefont {M.~D.}\ \bibnamefont
  {Hutchings}}, \bibinfo {author} {\bibfnamefont {J.~B.}\ \bibnamefont
  {Hertzberg}}, \bibinfo {author} {\bibfnamefont {Y.}~\bibnamefont {Liu}},
  \bibinfo {author} {\bibfnamefont {N.~T.}\ \bibnamefont {Bronn}}, \bibinfo
  {author} {\bibfnamefont {G.~A.}\ \bibnamefont {Keefe}}, \bibinfo {author}
  {\bibfnamefont {M.}~\bibnamefont {Brink}}, \bibinfo {author} {\bibfnamefont
  {J.~M.}\ \bibnamefont {Chow}}, \ and\ \bibinfo {author} {\bibfnamefont
  {B.~L.~T.}\ \bibnamefont {Plourde}},\ }\href {\doibase
  10.1103/PhysRevApplied.8.044003} {\bibfield  {journal} {\bibinfo  {journal}
  {Phys. Rev. Applied}\ }\textbf {\bibinfo {volume} {8}},\ \bibinfo {pages}
  {044003} (\bibinfo {year} {2017})}\BibitemShut {NoStop}%
\bibitem [{\citenamefont {You}\ \emph {et~al.}(2019)\citenamefont {You},
  \citenamefont {Sauls},\ and\ \citenamefont {Koch}}]{Koch19}%
  \BibitemOpen
  \bibfield  {author} {\bibinfo {author} {\bibfnamefont {X.}~\bibnamefont
  {You}}, \bibinfo {author} {\bibfnamefont {J.~A.}\ \bibnamefont {Sauls}}, \
  and\ \bibinfo {author} {\bibfnamefont {J.}~\bibnamefont {Koch}},\ }\href@noop
  {} {\enquote {\bibinfo {title} {Circuit quantization in the presence of
  time-dependent external flux},}\ } (\bibinfo {year} {2019}),\ \Eprint
  {http://arxiv.org/abs/arXiv:1902.04734} {arXiv:1902.04734} \BibitemShut
  {NoStop}%
\bibitem [{\citenamefont {Caldwell}\ \emph {et~al.}(2018)\citenamefont
  {Caldwell}, \citenamefont {Didier}, \citenamefont {Ryan}, \citenamefont
  {Sete}, \citenamefont {Hudson}, \citenamefont {Karalekas}, \citenamefont
  {Manenti}, \citenamefont {da~Silva}, \citenamefont {Sinclair}, \citenamefont
  {Acala}, \citenamefont {Alidoust}, \citenamefont {Angeles}, \citenamefont
  {Bestwick}, \citenamefont {Block}, \citenamefont {Bloom}, \citenamefont
  {Bradley}, \citenamefont {Bui}, \citenamefont {Capelluto}, \citenamefont
  {Chilcott}, \citenamefont {Cordova}, \citenamefont {Crossman}, \citenamefont
  {Curtis}, \citenamefont {Deshpande}, \citenamefont {Bouayadi}, \citenamefont
  {Girshovich}, \citenamefont {Hong}, \citenamefont {Kuang}, \citenamefont
  {Lenihan}, \citenamefont {Manning}, \citenamefont {Marchenkov}, \citenamefont
  {Marshall}, \citenamefont {Maydra}, \citenamefont {Mohan}, \citenamefont
  {O'Brien}, \citenamefont {Osborn}, \citenamefont {Otterbach}, \citenamefont
  {Papageorge}, \citenamefont {Paquette}, \citenamefont {Pelstring},
  \citenamefont {Polloreno}, \citenamefont {Prawiroatmodjo}, \citenamefont
  {Rawat}, \citenamefont {Reagor}, \citenamefont {Renzas}, \citenamefont
  {Rubin}, \citenamefont {Russell}, \citenamefont {Rust}, \citenamefont
  {Scarabelli}, \citenamefont {Scheer}, \citenamefont {Selvanayagam},
  \citenamefont {Smith}, \citenamefont {Staley}, \citenamefont {Suska},
  \citenamefont {Tezak}, \citenamefont {Thompson}, \citenamefont {To},
  \citenamefont {Vahidpour}, \citenamefont {Vodrahalli}, \citenamefont
  {Whyland}, \citenamefont {Yadav}, \citenamefont {Zeng},\ and\ \citenamefont
  {Rigetti}}]{Rigetti-blue_2017}%
  \BibitemOpen
  \bibfield  {author} {\bibinfo {author} {\bibfnamefont {S.~A.}\ \bibnamefont
  {Caldwell}}, \bibinfo {author} {\bibfnamefont {N.}~\bibnamefont {Didier}},
  \bibinfo {author} {\bibfnamefont {C.~A.}\ \bibnamefont {Ryan}}, \bibinfo
  {author} {\bibfnamefont {E.~A.}\ \bibnamefont {Sete}}, \bibinfo {author}
  {\bibfnamefont {A.}~\bibnamefont {Hudson}}, \bibinfo {author} {\bibfnamefont
  {P.}~\bibnamefont {Karalekas}}, \bibinfo {author} {\bibfnamefont
  {R.}~\bibnamefont {Manenti}}, \bibinfo {author} {\bibfnamefont {M.~P.}\
  \bibnamefont {da~Silva}}, \bibinfo {author} {\bibfnamefont {R.}~\bibnamefont
  {Sinclair}}, \bibinfo {author} {\bibfnamefont {E.}~\bibnamefont {Acala}},
  \bibinfo {author} {\bibfnamefont {N.}~\bibnamefont {Alidoust}}, \bibinfo
  {author} {\bibfnamefont {J.}~\bibnamefont {Angeles}}, \bibinfo {author}
  {\bibfnamefont {A.}~\bibnamefont {Bestwick}}, \bibinfo {author}
  {\bibfnamefont {M.}~\bibnamefont {Block}}, \bibinfo {author} {\bibfnamefont
  {B.}~\bibnamefont {Bloom}}, \bibinfo {author} {\bibfnamefont
  {A.}~\bibnamefont {Bradley}}, \bibinfo {author} {\bibfnamefont
  {C.}~\bibnamefont {Bui}}, \bibinfo {author} {\bibfnamefont {L.}~\bibnamefont
  {Capelluto}}, \bibinfo {author} {\bibfnamefont {R.}~\bibnamefont {Chilcott}},
  \bibinfo {author} {\bibfnamefont {J.}~\bibnamefont {Cordova}}, \bibinfo
  {author} {\bibfnamefont {G.}~\bibnamefont {Crossman}}, \bibinfo {author}
  {\bibfnamefont {M.}~\bibnamefont {Curtis}}, \bibinfo {author} {\bibfnamefont
  {S.}~\bibnamefont {Deshpande}}, \bibinfo {author} {\bibfnamefont {T.~E.}\
  \bibnamefont {Bouayadi}}, \bibinfo {author} {\bibfnamefont {D.}~\bibnamefont
  {Girshovich}}, \bibinfo {author} {\bibfnamefont {S.}~\bibnamefont {Hong}},
  \bibinfo {author} {\bibfnamefont {K.}~\bibnamefont {Kuang}}, \bibinfo
  {author} {\bibfnamefont {M.}~\bibnamefont {Lenihan}}, \bibinfo {author}
  {\bibfnamefont {T.}~\bibnamefont {Manning}}, \bibinfo {author} {\bibfnamefont
  {A.}~\bibnamefont {Marchenkov}}, \bibinfo {author} {\bibfnamefont
  {J.}~\bibnamefont {Marshall}}, \bibinfo {author} {\bibfnamefont
  {R.}~\bibnamefont {Maydra}}, \bibinfo {author} {\bibfnamefont
  {Y.}~\bibnamefont {Mohan}}, \bibinfo {author} {\bibfnamefont
  {W.}~\bibnamefont {O'Brien}}, \bibinfo {author} {\bibfnamefont
  {C.}~\bibnamefont {Osborn}}, \bibinfo {author} {\bibfnamefont
  {J.}~\bibnamefont {Otterbach}}, \bibinfo {author} {\bibfnamefont
  {A.}~\bibnamefont {Papageorge}}, \bibinfo {author} {\bibfnamefont {J.-P.}\
  \bibnamefont {Paquette}}, \bibinfo {author} {\bibfnamefont {M.}~\bibnamefont
  {Pelstring}}, \bibinfo {author} {\bibfnamefont {A.}~\bibnamefont
  {Polloreno}}, \bibinfo {author} {\bibfnamefont {G.}~\bibnamefont
  {Prawiroatmodjo}}, \bibinfo {author} {\bibfnamefont {V.}~\bibnamefont
  {Rawat}}, \bibinfo {author} {\bibfnamefont {M.}~\bibnamefont {Reagor}},
  \bibinfo {author} {\bibfnamefont {R.}~\bibnamefont {Renzas}}, \bibinfo
  {author} {\bibfnamefont {N.}~\bibnamefont {Rubin}}, \bibinfo {author}
  {\bibfnamefont {D.}~\bibnamefont {Russell}}, \bibinfo {author} {\bibfnamefont
  {M.}~\bibnamefont {Rust}}, \bibinfo {author} {\bibfnamefont {D.}~\bibnamefont
  {Scarabelli}}, \bibinfo {author} {\bibfnamefont {M.}~\bibnamefont {Scheer}},
  \bibinfo {author} {\bibfnamefont {M.}~\bibnamefont {Selvanayagam}}, \bibinfo
  {author} {\bibfnamefont {R.}~\bibnamefont {Smith}}, \bibinfo {author}
  {\bibfnamefont {A.}~\bibnamefont {Staley}}, \bibinfo {author} {\bibfnamefont
  {M.}~\bibnamefont {Suska}}, \bibinfo {author} {\bibfnamefont
  {N.}~\bibnamefont {Tezak}}, \bibinfo {author} {\bibfnamefont {D.~C.}\
  \bibnamefont {Thompson}}, \bibinfo {author} {\bibfnamefont {T.-W.}\
  \bibnamefont {To}}, \bibinfo {author} {\bibfnamefont {M.}~\bibnamefont
  {Vahidpour}}, \bibinfo {author} {\bibfnamefont {N.}~\bibnamefont
  {Vodrahalli}}, \bibinfo {author} {\bibfnamefont {T.}~\bibnamefont {Whyland}},
  \bibinfo {author} {\bibfnamefont {K.}~\bibnamefont {Yadav}}, \bibinfo
  {author} {\bibfnamefont {W.}~\bibnamefont {Zeng}}, \ and\ \bibinfo {author}
  {\bibfnamefont {C.}~\bibnamefont {Rigetti}},\ }\href {\doibase
  10.1103/PhysRevApplied.10.034050} {\bibfield  {journal} {\bibinfo  {journal}
  {Phys. Rev. Applied}\ }\textbf {\bibinfo {volume} {10}},\ \bibinfo {pages}
  {034050} (\bibinfo {year} {2018})}\BibitemShut {NoStop}%
\bibitem [{\citenamefont {Reagor}\ \emph {et~al.}(2018)\citenamefont {Reagor},
  \citenamefont {Osborn}, \citenamefont {Tezak}, \citenamefont {Staley},
  \citenamefont {Prawiroatmodjo}, \citenamefont {Scheer}, \citenamefont
  {Alidoust}, \citenamefont {Sete}, \citenamefont {Didier}, \citenamefont
  {da~Silva}, \citenamefont {Acala}, \citenamefont {Angeles}, \citenamefont
  {Bestwick}, \citenamefont {Block}, \citenamefont {Bloom}, \citenamefont
  {Bradley}, \citenamefont {Bui}, \citenamefont {Caldwell}, \citenamefont
  {Capelluto}, \citenamefont {Chilcott}, \citenamefont {Cordova}, \citenamefont
  {Crossman}, \citenamefont {Curtis}, \citenamefont {Deshpande}, \citenamefont
  {El~Bouayadi}, \citenamefont {Girshovich}, \citenamefont {Hong},
  \citenamefont {Hudson}, \citenamefont {Karalekas}, \citenamefont {Kuang},
  \citenamefont {Lenihan}, \citenamefont {Manenti}, \citenamefont {Manning},
  \citenamefont {Marshall}, \citenamefont {Mohan}, \citenamefont
  {O{\textquoteright}Brien}, \citenamefont {Otterbach}, \citenamefont
  {Papageorge}, \citenamefont {Paquette}, \citenamefont {Pelstring},
  \citenamefont {Polloreno}, \citenamefont {Rawat}, \citenamefont {Ryan},
  \citenamefont {Renzas}, \citenamefont {Rubin}, \citenamefont {Russel},
  \citenamefont {Rust}, \citenamefont {Scarabelli}, \citenamefont
  {Selvanayagam}, \citenamefont {Sinclair}, \citenamefont {Smith},
  \citenamefont {Suska}, \citenamefont {To}, \citenamefont {Vahidpour},
  \citenamefont {Vodrahalli}, \citenamefont {Whyland}, \citenamefont {Yadav},
  \citenamefont {Zeng},\ and\ \citenamefont {Rigetti}}]{Rigetti-white_2017}%
  \BibitemOpen
  \bibfield  {author} {\bibinfo {author} {\bibfnamefont {M.}~\bibnamefont
  {Reagor}}, \bibinfo {author} {\bibfnamefont {C.~B.}\ \bibnamefont {Osborn}},
  \bibinfo {author} {\bibfnamefont {N.}~\bibnamefont {Tezak}}, \bibinfo
  {author} {\bibfnamefont {A.}~\bibnamefont {Staley}}, \bibinfo {author}
  {\bibfnamefont {G.}~\bibnamefont {Prawiroatmodjo}}, \bibinfo {author}
  {\bibfnamefont {M.}~\bibnamefont {Scheer}}, \bibinfo {author} {\bibfnamefont
  {N.}~\bibnamefont {Alidoust}}, \bibinfo {author} {\bibfnamefont {E.~A.}\
  \bibnamefont {Sete}}, \bibinfo {author} {\bibfnamefont {N.}~\bibnamefont
  {Didier}}, \bibinfo {author} {\bibfnamefont {M.~P.}\ \bibnamefont
  {da~Silva}}, \bibinfo {author} {\bibfnamefont {E.}~\bibnamefont {Acala}},
  \bibinfo {author} {\bibfnamefont {J.}~\bibnamefont {Angeles}}, \bibinfo
  {author} {\bibfnamefont {A.}~\bibnamefont {Bestwick}}, \bibinfo {author}
  {\bibfnamefont {M.}~\bibnamefont {Block}}, \bibinfo {author} {\bibfnamefont
  {B.}~\bibnamefont {Bloom}}, \bibinfo {author} {\bibfnamefont
  {A.}~\bibnamefont {Bradley}}, \bibinfo {author} {\bibfnamefont
  {C.}~\bibnamefont {Bui}}, \bibinfo {author} {\bibfnamefont {S.}~\bibnamefont
  {Caldwell}}, \bibinfo {author} {\bibfnamefont {L.}~\bibnamefont {Capelluto}},
  \bibinfo {author} {\bibfnamefont {R.}~\bibnamefont {Chilcott}}, \bibinfo
  {author} {\bibfnamefont {J.}~\bibnamefont {Cordova}}, \bibinfo {author}
  {\bibfnamefont {G.}~\bibnamefont {Crossman}}, \bibinfo {author}
  {\bibfnamefont {M.}~\bibnamefont {Curtis}}, \bibinfo {author} {\bibfnamefont
  {S.}~\bibnamefont {Deshpande}}, \bibinfo {author} {\bibfnamefont
  {T.}~\bibnamefont {El~Bouayadi}}, \bibinfo {author} {\bibfnamefont
  {D.}~\bibnamefont {Girshovich}}, \bibinfo {author} {\bibfnamefont
  {S.}~\bibnamefont {Hong}}, \bibinfo {author} {\bibfnamefont {A.}~\bibnamefont
  {Hudson}}, \bibinfo {author} {\bibfnamefont {P.}~\bibnamefont {Karalekas}},
  \bibinfo {author} {\bibfnamefont {K.}~\bibnamefont {Kuang}}, \bibinfo
  {author} {\bibfnamefont {M.}~\bibnamefont {Lenihan}}, \bibinfo {author}
  {\bibfnamefont {R.}~\bibnamefont {Manenti}}, \bibinfo {author} {\bibfnamefont
  {T.}~\bibnamefont {Manning}}, \bibinfo {author} {\bibfnamefont
  {J.}~\bibnamefont {Marshall}}, \bibinfo {author} {\bibfnamefont
  {Y.}~\bibnamefont {Mohan}}, \bibinfo {author} {\bibfnamefont
  {W.}~\bibnamefont {O{\textquoteright}Brien}}, \bibinfo {author}
  {\bibfnamefont {J.}~\bibnamefont {Otterbach}}, \bibinfo {author}
  {\bibfnamefont {A.}~\bibnamefont {Papageorge}}, \bibinfo {author}
  {\bibfnamefont {J.-P.}\ \bibnamefont {Paquette}}, \bibinfo {author}
  {\bibfnamefont {M.}~\bibnamefont {Pelstring}}, \bibinfo {author}
  {\bibfnamefont {A.}~\bibnamefont {Polloreno}}, \bibinfo {author}
  {\bibfnamefont {V.}~\bibnamefont {Rawat}}, \bibinfo {author} {\bibfnamefont
  {C.~A.}\ \bibnamefont {Ryan}}, \bibinfo {author} {\bibfnamefont
  {R.}~\bibnamefont {Renzas}}, \bibinfo {author} {\bibfnamefont
  {N.}~\bibnamefont {Rubin}}, \bibinfo {author} {\bibfnamefont
  {D.}~\bibnamefont {Russel}}, \bibinfo {author} {\bibfnamefont
  {M.}~\bibnamefont {Rust}}, \bibinfo {author} {\bibfnamefont {D.}~\bibnamefont
  {Scarabelli}}, \bibinfo {author} {\bibfnamefont {M.}~\bibnamefont
  {Selvanayagam}}, \bibinfo {author} {\bibfnamefont {R.}~\bibnamefont
  {Sinclair}}, \bibinfo {author} {\bibfnamefont {R.}~\bibnamefont {Smith}},
  \bibinfo {author} {\bibfnamefont {M.}~\bibnamefont {Suska}}, \bibinfo
  {author} {\bibfnamefont {T.-W.}\ \bibnamefont {To}}, \bibinfo {author}
  {\bibfnamefont {M.}~\bibnamefont {Vahidpour}}, \bibinfo {author}
  {\bibfnamefont {N.}~\bibnamefont {Vodrahalli}}, \bibinfo {author}
  {\bibfnamefont {T.}~\bibnamefont {Whyland}}, \bibinfo {author} {\bibfnamefont
  {K.}~\bibnamefont {Yadav}}, \bibinfo {author} {\bibfnamefont
  {W.}~\bibnamefont {Zeng}}, \ and\ \bibinfo {author} {\bibfnamefont {C.~T.}\
  \bibnamefont {Rigetti}},\ }\href {\doibase 10.1126/sciadv.aao3603} {\bibfield
   {journal} {\bibinfo  {journal} {Science Advances}\ }\textbf {\bibinfo
  {volume} {4}} (\bibinfo {year} {2018}),\ 10.1126/sciadv.aao3603}\BibitemShut
  {NoStop}%
\bibitem [{\citenamefont {Didier}\ \emph {et~al.}(2015)\citenamefont {Didier},
  \citenamefont {Bourassa},\ and\ \citenamefont {Blais}}]{Didier15}%
  \BibitemOpen
  \bibfield  {author} {\bibinfo {author} {\bibfnamefont {N.}~\bibnamefont
  {Didier}}, \bibinfo {author} {\bibfnamefont {J.}~\bibnamefont {Bourassa}}, \
  and\ \bibinfo {author} {\bibfnamefont {A.}~\bibnamefont {Blais}},\ }\href
  {\doibase 10.1103/PhysRevLett.115.203601} {\bibfield  {journal} {\bibinfo
  {journal} {Phys. Rev. Lett.}\ }\textbf {\bibinfo {volume} {115}},\ \bibinfo
  {pages} {203601} (\bibinfo {year} {2015})}\BibitemShut {NoStop}%
\bibitem [{Note1()}]{Note1}%
  \BibitemOpen
  \bibinfo {note} {Similar dynamical sweet spots have been previously described
  for charge noise in superconducting qubits~\cite
  {Sillanpaa2012}.}\BibitemShut {Stop}%
\bibitem [{\citenamefont {Hong}\ \emph {et~al.}(2019)\citenamefont {Hong},
  \citenamefont {Papageorge}, \citenamefont {Sivarajah}, \citenamefont
  {Crossman}, \citenamefont {Dider}, \citenamefont {Polloreno}, \citenamefont
  {Sete}, \citenamefont {Turkowski}, \citenamefont {da~Silva},\ and\
  \citenamefont {Johnson}}]{Hong19}%
  \BibitemOpen
  \bibfield  {author} {\bibinfo {author} {\bibfnamefont {S.~S.}\ \bibnamefont
  {Hong}}, \bibinfo {author} {\bibfnamefont {A.~T.}\ \bibnamefont
  {Papageorge}}, \bibinfo {author} {\bibfnamefont {P.}~\bibnamefont
  {Sivarajah}}, \bibinfo {author} {\bibfnamefont {G.}~\bibnamefont {Crossman}},
  \bibinfo {author} {\bibfnamefont {N.}~\bibnamefont {Dider}}, \bibinfo
  {author} {\bibfnamefont {A.~M.}\ \bibnamefont {Polloreno}}, \bibinfo {author}
  {\bibfnamefont {E.~A.}\ \bibnamefont {Sete}}, \bibinfo {author}
  {\bibfnamefont {S.~W.}\ \bibnamefont {Turkowski}}, \bibinfo {author}
  {\bibfnamefont {M.~P.}\ \bibnamefont {da~Silva}}, \ and\ \bibinfo {author}
  {\bibfnamefont {B.~R.}\ \bibnamefont {Johnson}},\ }\href@noop {} {\enquote
  {\bibinfo {title} {Demonstration of a parametrically-activated entangling
  gate protected from flux noise},}\ } (\bibinfo {year} {2019}),\ \Eprint
  {http://arxiv.org/abs/arXiv:1901.08035} {arXiv:1901.08035} \BibitemShut
  {NoStop}%
\bibitem [{\citenamefont {Sete}\ \emph {et~al.}(2017)\citenamefont {Sete},
  \citenamefont {Reagor}, \citenamefont {Didier},\ and\ \citenamefont
  {Rigetti}}]{Sete2017}%
  \BibitemOpen
  \bibfield  {author} {\bibinfo {author} {\bibfnamefont {E.~A.}\ \bibnamefont
  {Sete}}, \bibinfo {author} {\bibfnamefont {M.~J.}\ \bibnamefont {Reagor}},
  \bibinfo {author} {\bibfnamefont {N.}~\bibnamefont {Didier}}, \ and\ \bibinfo
  {author} {\bibfnamefont {C.~T.}\ \bibnamefont {Rigetti}},\ }\href {\doibase
  10.1103/PhysRevApplied.8.024004} {\bibfield  {journal} {\bibinfo  {journal}
  {Phys. Rev. Applied}\ }\textbf {\bibinfo {volume} {8}},\ \bibinfo {pages}
  {024004} (\bibinfo {year} {2017})}\BibitemShut {NoStop}%
\bibitem [{\citenamefont {Paladino}\ \emph {et~al.}(2014)\citenamefont
  {Paladino}, \citenamefont {Galperin}, \citenamefont {Falci},\ and\
  \citenamefont {Altshuler}}]{PGFA14}%
  \BibitemOpen
  \bibfield  {author} {\bibinfo {author} {\bibfnamefont {E.}~\bibnamefont
  {Paladino}}, \bibinfo {author} {\bibfnamefont {Y.~M.}\ \bibnamefont
  {Galperin}}, \bibinfo {author} {\bibfnamefont {G.}~\bibnamefont {Falci}}, \
  and\ \bibinfo {author} {\bibfnamefont {B.~L.}\ \bibnamefont {Altshuler}},\
  }\href {\doibase 10.1103/RevModPhys.86.361} {\bibfield  {journal} {\bibinfo
  {journal} {Rev. Mod. Phys.}\ }\textbf {\bibinfo {volume} {86}},\ \bibinfo
  {pages} {361} (\bibinfo {year} {2014})}\BibitemShut {NoStop}%
\bibitem [{\citenamefont {Didier}\ \emph {et~al.}(2018)\citenamefont {Didier},
  \citenamefont {Sete}, \citenamefont {da~Silva},\ and\ \citenamefont
  {Rigetti}}]{Nico17}%
  \BibitemOpen
  \bibfield  {author} {\bibinfo {author} {\bibfnamefont {N.}~\bibnamefont
  {Didier}}, \bibinfo {author} {\bibfnamefont {E.~A.}\ \bibnamefont {Sete}},
  \bibinfo {author} {\bibfnamefont {M.~P.}\ \bibnamefont {da~Silva}}, \ and\
  \bibinfo {author} {\bibfnamefont {C.}~\bibnamefont {Rigetti}},\ }\href
  {\doibase 10.1103/PhysRevA.97.022330} {\bibfield  {journal} {\bibinfo
  {journal} {Phys. Rev. A}\ }\textbf {\bibinfo {volume} {97}},\ \bibinfo
  {pages} {022330} (\bibinfo {year} {2018})}\BibitemShut {NoStop}%
\bibitem [{Not()}]{Note}%
  \BibitemOpen
  \href@noop {} {\ }\bibinfo {note} {We have used the method described in
  Astron. Astrophys. 300, 707 (1995) to generate the numerical 1/f and white
  noise time traces.}\BibitemShut {Stop}%
\bibitem [{\citenamefont {Purcell}(1946)}]{Purcell}%
  \BibitemOpen
  \bibfield  {author} {\bibinfo {author} {\bibfnamefont {E.~M.}\ \bibnamefont
  {Purcell}},\ }\href@noop {} {\bibfield  {journal} {\bibinfo  {journal} {Phys.
  Rev.}\ }\textbf {\bibinfo {volume} {69}},\ \bibinfo {pages} {681} (\bibinfo
  {year} {1946})}\BibitemShut {NoStop}%
\bibitem [{\citenamefont {Houck}\ \emph {et~al.}(2008)\citenamefont {Houck},
  \citenamefont {Schreier}, \citenamefont {Johnson}, \citenamefont {Chow},
  \citenamefont {Koch}, \citenamefont {Gambetta}, \citenamefont {Schuster},
  \citenamefont {Frunzio}, \citenamefont {Devoret}, \citenamefont {Girvin},\
  and\ \citenamefont {Schoelkopf}}]{Houck2008}%
  \BibitemOpen
  \bibfield  {author} {\bibinfo {author} {\bibfnamefont {A.~A.}\ \bibnamefont
  {Houck}}, \bibinfo {author} {\bibfnamefont {J.~A.}\ \bibnamefont {Schreier}},
  \bibinfo {author} {\bibfnamefont {B.~R.}\ \bibnamefont {Johnson}}, \bibinfo
  {author} {\bibfnamefont {J.~M.}\ \bibnamefont {Chow}}, \bibinfo {author}
  {\bibfnamefont {J.}~\bibnamefont {Koch}}, \bibinfo {author} {\bibfnamefont
  {J.~M.}\ \bibnamefont {Gambetta}}, \bibinfo {author} {\bibfnamefont {D.~I.}\
  \bibnamefont {Schuster}}, \bibinfo {author} {\bibfnamefont {L.}~\bibnamefont
  {Frunzio}}, \bibinfo {author} {\bibfnamefont {M.~H.}\ \bibnamefont
  {Devoret}}, \bibinfo {author} {\bibfnamefont {S.~M.}\ \bibnamefont {Girvin}},
  \ and\ \bibinfo {author} {\bibfnamefont {R.~J.}\ \bibnamefont {Schoelkopf}},\
  }\href {\doibase 10.1103/PhysRevLett.101.080502} {\bibfield  {journal}
  {\bibinfo  {journal} {Phys. Rev. Lett.}\ }\textbf {\bibinfo {volume} {101}},\
  \bibinfo {pages} {080502} (\bibinfo {year} {2008})}\BibitemShut {NoStop}%
\bibitem [{\citenamefont {Sete}\ \emph {et~al.}(2014)\citenamefont {Sete},
  \citenamefont {Gambetta},\ and\ \citenamefont {Korotkov}}]{Sete2014}%
  \BibitemOpen
  \bibfield  {author} {\bibinfo {author} {\bibfnamefont {E.~A.}\ \bibnamefont
  {Sete}}, \bibinfo {author} {\bibfnamefont {J.~M.}\ \bibnamefont {Gambetta}},
  \ and\ \bibinfo {author} {\bibfnamefont {A.~N.}\ \bibnamefont {Korotkov}},\
  }\href {\doibase 10.1103/PhysRevB.89.104516} {\bibfield  {journal} {\bibinfo
  {journal} {Phys. Rev. B}\ }\textbf {\bibinfo {volume} {89}},\ \bibinfo
  {pages} {104516} (\bibinfo {year} {2014})}\BibitemShut {NoStop}%
\bibitem [{\citenamefont {Rol}\ \emph {et~al.}(2019)\citenamefont {Rol},
  \citenamefont {Battistel}, \citenamefont {Malinowski}, \citenamefont
  {Bultink}, \citenamefont {Tarasinski}, \citenamefont {Vollmer}, \citenamefont
  {Haider}, \citenamefont {Muthusubramanian}, \citenamefont {Bruno},
  \citenamefont {Terhal},\ and\ \citenamefont {DiCarlo}}]{DiCarlo19}%
  \BibitemOpen
  \bibfield  {author} {\bibinfo {author} {\bibfnamefont {M.~A.}\ \bibnamefont
  {Rol}}, \bibinfo {author} {\bibfnamefont {F.}~\bibnamefont {Battistel}},
  \bibinfo {author} {\bibfnamefont {F.~K.}\ \bibnamefont {Malinowski}},
  \bibinfo {author} {\bibfnamefont {C.~C.}\ \bibnamefont {Bultink}}, \bibinfo
  {author} {\bibfnamefont {B.~M.}\ \bibnamefont {Tarasinski}}, \bibinfo
  {author} {\bibfnamefont {R.}~\bibnamefont {Vollmer}}, \bibinfo {author}
  {\bibfnamefont {N.}~\bibnamefont {Haider}}, \bibinfo {author} {\bibfnamefont
  {N.}~\bibnamefont {Muthusubramanian}}, \bibinfo {author} {\bibfnamefont
  {A.}~\bibnamefont {Bruno}}, \bibinfo {author} {\bibfnamefont {B.~M.}\
  \bibnamefont {Terhal}}, \ and\ \bibinfo {author} {\bibfnamefont
  {L.}~\bibnamefont {DiCarlo}},\ }\href@noop {} {\  (\bibinfo {year} {2019})},\
  \Eprint {http://arxiv.org/abs/arXiv:1903.02492} {arXiv:1903.02492}
  \BibitemShut {NoStop}%
\bibitem [{\citenamefont {Nielsen}(2002)}]{Nielsen2002}%
  \BibitemOpen
  \bibfield  {author} {\bibinfo {author} {\bibfnamefont {M.~A.}\ \bibnamefont
  {Nielsen}},\ }\href {\doibase https://doi.org/10.1016/S0375-9601(02)01272-0}
  {\bibfield  {journal} {\bibinfo  {journal} {Physics Letters A}\ }\textbf
  {\bibinfo {volume} {303}},\ \bibinfo {pages} {249 } (\bibinfo {year}
  {2002})}\BibitemShut {NoStop}%
\bibitem [{\citenamefont {Pirkkalainen}\ \emph {et~al.}(2012)\citenamefont
  {Pirkkalainen}, \citenamefont {Cho}, \citenamefont {Li}, \citenamefont
  {Paraoanu}, \citenamefont {Hakonen},\ and\ \citenamefont
  {Sillanpaa}}]{Sillanpaa2012}%
  \BibitemOpen
  \bibfield  {author} {\bibinfo {author} {\bibfnamefont {J.~M.}\ \bibnamefont
  {Pirkkalainen}}, \bibinfo {author} {\bibfnamefont {S.~U.}\ \bibnamefont
  {Cho}}, \bibinfo {author} {\bibfnamefont {J.}~\bibnamefont {Li}}, \bibinfo
  {author} {\bibfnamefont {G.~S.}\ \bibnamefont {Paraoanu}}, \bibinfo {author}
  {\bibfnamefont {P.~J.}\ \bibnamefont {Hakonen}}, \ and\ \bibinfo {author}
  {\bibfnamefont {M.~A.}\ \bibnamefont {Sillanpaa}},\ }\href {\doibase
  10.1038/nature11821} {\  (\bibinfo {year} {2012}),\ 10.1038/nature11821},\
  \Eprint {http://arxiv.org/abs/arXiv:1207.1637} {arXiv:1207.1637} \BibitemShut
  {NoStop}%
\end{thebibliography}
\end{document}